\documentclass[12pt,a4paper]{article}
\pdfoutput=1
\usepackage{graphicx}
\usepackage{appendix}
\usepackage{amsmath,amsfonts,amssymb,setspace}
\usepackage{xcolor}
\usepackage{comment}
\newcommand{\e}{\epsilon}
\newcommand{\be}[1]{\begin{equation}\label{#1} }
\newcommand{\ee}{\end{equation}}
\newcommand{\bea}[1]{\begin{eqnarray}\label{#1} }
\newcommand{\eea}{\end{eqnarray}}
\newcommand{\p}{\partial}

\newcommand{\bL}{\bar{{\mathcal{L}}}}

\renewcommand{\>}{\rangle}

\newcommand{\za}{|0\rangle_a}
\newcommand{\zc}{|0\rangle_c}

\newcommand{\C}{\tilde{C}}

\newcommand{\s}{\sigma}

\usepackage{caption}
\usepackage{subcaption}
\usepackage{hyperref}
\usepackage{cite}
\usepackage{tocloft}
\usepackage[margin=0.75in]{geometry}
\hypersetup{colorlinks=false, linkcolor=blue, citecolor=red}
	\def\be{\begin{equation}}
	\def\ee{\end{equation}}

	\newcommand{\vs}[1]{\vspace{#1 mm}}
	
\begin{document}

	\begin{flushright}
		%
	\end{flushright}
	\begin{center}
		{\Large{\bf From CFTs to theories with Bondi-Metzner-Sachs symmetries: Complexity
    and out-of-time-ordered correlators}}\\

\vs{10}

{\large
Aritra Banerjee${}^{a,\,}$\footnote{\url{aritra.banerjee@oist.jp}}, Arpan Bhattacharyya${}^{b,\,}$\footnote{\url{abhattacharyya@iitgn.ac.in}}, \\ Priya Drashni${}^{b,\,}$\footnote{\url{priyad@iitgn.ac.in}}, Srinidhi Pawar${}^{b,\,}$\footnote{\url{srinidhip@iitgn.ac.in }}}

\vskip 0.3in

{\it ${}^{a}$ Okinawa Institute of Science and Technology, \\1919-1 Tancha, Onna-son, Okinawa 904-0495, Japan }\vskip .5mm

{\it ${}^{b}$Indian Institute of Technology, Gandhinagar,\\Gujarat-382355, India}\vskip .5mm

\vskip.5mm

\end{center}

\vskip 0.35in

\begin{abstract}
We probe the contraction from $2d$ relativistic CFTs to theories with Bondi-Metzner-Sachs (BMS) symmetries, or equivalently Conformal Carroll symmetries, using diagnostics of quantum chaos. Starting from an Ultrarelativistic limit on a relativistic scalar field theory and following through at the quantum level using an oscillator representation of states, one can show the CFT$_2$ vacuum evolves smoothly into a BMS$_3$ vacuum in the form of a squeezed state. Computing circuit complexity of this transmutation using the covariance matrix approach shows clear divergences when the BMS point is hit or equivalently when the target state becomes a boundary state. We also find similar behaviour of the circuit complexity calculated from methods of information geometry. Furthermore, we discuss the hamiltonian evolution of the system and investigate Out-of-time-ordered correlators (OTOCs) and operator growth complexity, both of which turn out to scale polynomially with time at the BMS point.

\end{abstract}

	\newpage

\tableofcontents

\section{Introduction}
The holy grail of quantum gravity is one of the most sought after treasures in modern theoretical physics, and many avenues to achieve that exist, with String Theory being arguably the most successful one. A very fruitful effort towards the same has taken shape over the past two decades, famously known as the holographic duality, whose most well-known avatar is the AdS/CFT correspondence  \cite{Maldacena:1997re}. The AdS/CFT correspondence states that gravity in asymptotically Anti-de Sitter (AdS) spacetime is equivalent to a quantum field theory with conformal invariance living on the boundary of the AdS space and observables calculated from either theory can be matched up using this equivalence. The advent of AdS/CFT has given rise to various multidisciplinary research fields involving many seemingly disconnected branches of Physics, including string theory, Black Hole physics, condensed matter physics,  and, more recently, quantum information theory.

However, our visible Universe seems to have nothing to do with AdS, as experimental signatures argue against a negative value of the cosmological constant. An extension of this duality to de Sitter space (dS) has not been satisfactorily formulated yet. Hence, the prospect of extending the duality to asymptotically Flat spacetime seems exciting enough to pursue. We recall that a central idea in the so-called holographic dictionary is the asymptotic symmetry of the bulk $(d+1)$ dimensional gravity theory agreeing precisely with the global symmetry of the dual field theory living in one lower dimension, and both are the conformal symmetry in $d$ dimensions. The asymptotic symmetry for four-dimensional flat Minkowski spacetime containing Einstein gravity, first studied by Bondi-van der Burg-Metzner-Sachs (BMS)\cite{Bondi:1,Sachs:1962zza}, is the so-called BMS symmetry. Hence the putative dual theory living on the boundary of flat spacetime, via this `Flat Holography', should also be invariant under the same BMS symmetries, which is the main message of the whole program. Consequently, many investigations on BMS invariant field theories (BMSFTs) in various dimensions have appeared in the literature over the last few years \cite{Bagchi:2012yk,Bagchi:2012xr,Barnich:2012xq,Bagchi:2013qva, Bagchi:2015wna, Bagchi:2014iea,Jiang:2017ecm,Hijano:2017eii, Bagchi:2013lma, Detournay:2014fva, Hartong:2015usd, Hartong:2015xda, Bagchi:2016geg, Barnich:2014cwa, Fareghbal:2014qga, Grumiller:2019xna, Ciambelli:2018wre}. 

In another development related to Flat Holography, known as `Celestial Holography', the idea of holography for asymptotically flat spacetimes has been formulated as a correspondence between gravity on $4d$ flat space and a $2d$ CFT living on the celestial sphere. This way of thinking has attracted a lot of attention in the last few years in terms of linking asymptotic symmetries and scattering amplitudes and one can have a look at the excellent reviews \cite{Strominger:2017zoo, Pasterski:2021rjz, Raclariu:2021zjz} for a detailed understanding of these exciting structures.  What is more intriguing, there seems to be a very recently discovered reconciliation between these two seemingly different avenues put forward in \cite{Bagchi:2022emh}. This nicely packages ideas from Celestial Holography in the language of correlation functions of  Carrollian CFT (equivalently, BMSFTs) and links them naturally to $4d$ scattering amplitudes. For further intriguing ideas linking the two approaches, see also \cite{Donnay:2022aba}. 

In lower dimensions, especially in the case of three, the BMS group has a particularly simple structure and is isomorphic to the Galilean Conformal Algebra in two dimensions (GCA$_2$) \cite{Bagchi:2009my}. While the GCA$_2$ can be obtained from the non-relativistic (NR) contraction of the two-dimensional conformal algebra, which is two copies of the Virasoro algebra, the BMS$_3$ algebra is the ultra-relativistic (UR) contraction of the same, and can be shown to be isomorphic to a Conformal Carrollian Algebra in two dimensions (CCA$_2$) \cite{Duval_2014, Duval:2014uoa, Duval:2014uva}. All of this turns out to be a part of the general equivalence between BMS algebra in $(d+1)$ dimensions and Conformal Carrollian Algebra in $d$ dimensions. Carrollian symmetries occur whenever we encounter a null surface and Riemannian structures degenerate \cite{Barnich:2010eb,Bagchi:2010zz,Bagchi:2012cy} due to the closure of lightcones. Two-dimensional field theories with these symmetries are a very active research area, and various results have been obtained \cite{Ciambelli:2018ojf,Gupta:2020dtl,Bagchi:2022eav}. Despite the thorough investigations into Entanglement Entropy of such theories \cite{Bagchi:2014iea,Jiang:2017ecm,Hijano:2017eii,Donnay:2019jiz}, many other information-theoretic structures of this theory remain in the shadows. However, recently some more investigations into the quantum chaotic structure in BMSFTs have been detailed in \cite{Bagchi:2021qfe}. In this work, we will be trying to shed light on some unexplored issues, especially how certain information-theoretic markers change as a physical system goes through the contraction of conformal symmetries into BMS symmetries. Our focus will be on a free scalar field theory, which has appeared in the literature in various guises, including in the study of  null string theories \cite{Schild:1976vq,Isberg:1993av,Bagchi:2013bga,Bagchi:2015nca}, as a BMSFT action \cite{Hao:2021urq}, and as deformations of $2d$ CFT actions \cite{Rodriguez:2021tcz, Bagchi:2022nvj}. 

In quantum information theory, Quantum Circuit Complexity is a very useful tool to probe into the structure of an inherently quantum theory. The idea of complexity in  Quantum Information theory is simple. Given a suitable basis, it is a quantity that determines the minimum number of operations needed to perform the desired task. Specifically for a quantum system, the notion of complexity is associated with an efficient quantum circuit that takes a reference state (usually a state that can be prepared relatively easily in the `lab') into the desired target state given a set of quantum gates. In recent times, the notion of complexity has appeared extensively in the context of holography. In the context of AdS/CFT, certain geometrical objects have been interpreted as gravity dual of the circuit complexity of the dual field theory state. These proposals go by the names of \textit{complexity = volume} \cite{Stanford:2014jda}(maximal volume of co-dimension one bulk slice) and \textit{complexity = action}\cite{Brown:2015bva} (gravitational action defined on a certain Wheeler-De Witt patch inside the bulk space-time). This has spurred  lots of studies of circuit complexity in the context of quantum field theory \cite{Jefferson1,Chapman:2017rqy,Caputa:2017yrh,me1,Bhattacharyya:2018bbv,Hackl:2018ptj,Khan:2018rzm,Alves:2018qfv,Camargo:2018eof,Ali:2018aon,Bhattacharyya:2018wym,Caputa:2018kdj,Bhattacharyya:2019kvj,Flory:2020eot,Erdmenger:2020sup, cosmology1,cosmology2,DiGiulio:2020hlz,Caceres:2019pgf,Chen:2020nlj,Czech:2017ryf,Camargo:2019isp,Chapman:2018hou,Doroudiani:2019llj,Geng:2019yxo,Couch:2021wsm,Bhattacharyya:2021fii,Bhattacharyya:2019txx,Erdmenger:2021wzc,Chagnet:2021uvi,Bhattacharyya:2022ren} \footnote{This list is by no means exhaustive. Readers are referred to the reviews \cite{Chapman:2021jbh,Bhattacharyya:2021cwf,Koch:2021tvp}, and references therein for more details. Also, there are several other proposals for holographic complexity e.g. the ones discussed in \cite{Couch:2018phr,Belin:2021bga,Agon:2018zso,Belin:2018bpg}. Again this list is also not exhaustive at all.}.However, the ramifications of such constructions are far from well explored. Several methods of quantifying complexity in a QFT exist, and they all have their own advantages, see \cite{me1} for a detailed discussion. 

It has recently been proposed \cite{Bhattacharyya:2018bbv,Bhattacharyya:2019kvj}, that the circuit complexity can be used as a useful probe of flows between different quantum field theories (more specifically, as a probe of renormalization group flow) and quantum phase transitions. Motivated by this, in this paper, we will use circuit complexity to probe the purported ``flow'' from CFT to BMS invariant theories \cite{Bagchi:2022nvj}. 
Besides exploring circuit complexity, we will also discuss the Hamiltonian dynamics of the system. Intriguing new research has unearthed that quantum chaos in quantum many-body systems plays an important role in understanding some of the important open questions, e.g. thermalization, transport in quantum many-body systems, black hole information loss etc. \cite{NHJ,Jahnke:2018off}. In this paper, we will also compute  Out-Of-Time-Ordered correlators (OTOCs) for our system. It has been shown \cite{Kitaev2015,Larkin1969,Maldacena_2016}\footnote{For computation of OTOC in quantum mechanical systems interested readers are referred to \cite{Hashimoto:2017oit}.} that OTOC gives pertinent information about the Lyapunov exponent and the scrambling time \footnote{For a detailed review one can look at \cite{Jahnke:2018off} and the references therein.}. We will also study the nature of operator evolution in the Heisenberg picture for such a flow from CFT to BMS. The complexity associated to this process has been termed Krylov Complexity in the literature, and has been examined thoroughly \cite{Parker:2018yvk,Dymarsky:2021bjq,Barbon:2019wsy,Rabinovici:2020ryf,Kar:2021nbm,Caputa:2021sib,Balasubramanian:2022tpr, Bhattacharjee:2022vlt, Muck:2022xfc} \footnote{This list again does not do justice to the literature that has discussed related topics in recent times. Interested readers are referred to the references and citations of these papers.} in recent times. 
\par

The organization of the paper is as follows. In section (\ref{sec1}), we discuss the underlying model based on the massless scalar field and the limiting procedure to obtain the BMS vacuum from the CFT vacuum. In section (\ref{sec2}), we explore circuit complexity as a function of the contraction parameter from CFT to BMS. We observe that the complexity becomes divergent when the system hits the BMS point. To get further intuition about this diverging complexity, we study the associated information geometry in the section (\ref{sec3}). Specifically, we study the Fubini-study metric and the geodesic that connects the CFT and BMS vacuum on the state manifold. We also comment on the Berry Curvature for this process. Lastly, in the section (\ref{sec4}), we study the Hamiltonian of the system and the behaviour of the OTOCs. We find that the OTOC is a polynomial function of time in the BMS limit. A close study of the Krylov Complexity finds a similar polynomial scaling associated with operator evolution in this limit. We conclude in the section (\ref{sec5}) by summarizing our results and proposing future directions.

\section{Revisiting BMS$_3$ invariant scalar field} \label{sec1}
\subsection{The intrinsic model}
As discussed in the introduction, our core model concerns an Inönü-Wigner contraction from $2d$ relativistic conformal field theories to theories with BMS$_3$ as their symmetry algebra. A very well-studied example of this appears in the study of Null or Tensionless string theories  \cite{Schild:1976vq,Isberg:1993av,Bagchi:2013bga,Bagchi:2015nca}. In this limit, the worldsheet of the string becomes null, endowed with a degenerate metric and acquires a Carrollian structure, where the residual symmetry algebra coincides with that of BMS$_3$. From a CFT point of view, BMSFTs generically occur as a limit of the $2d$ conformal algebra, which is isomorphic to two copies of the Virasoro algebra. 
For completeness, these Virasoro generators on a cylinder parameterized by $(\sigma \sim \sigma+2\pi,\tau)$ is given by the following vector fields,
\begin{align}
    \mathcal{L}_{k}=\frac{i}{2}e^{i\,k\,(\tau+\sigma)}(\partial_{\tau}+\partial_{\s}),~~~\bar{\mathcal{L}}_{k}=\frac{i}{2}e^{i\,k\,(\tau-\sigma)}(\partial_{\tau}-\partial_{\s}).
\end{align}
At the level of mode expansions, these correspond to two independent sets of oscillators corresponding to holomorphic and anti-holomorphic sectors in the CFT. These generators satisfy the classical part of the Virasoro algebra
\begin{align}
     [\mathcal{L}_{k},\mathcal{L}_{k'}]=(k-k')\mathcal{L}_{k+k'},~~~ [\bar{\mathcal{L}}_{k},\bar{\mathcal{L}}_{k'}]=(k-k')\bar{\mathcal{L}}_{k+k'},
\end{align}
where one can add Virasoro central charges $c, \bar{c}$ to the algebra when quantized. Given the two Virasoro generators $\mathcal{L}_k$ and $\bL_k$, the  contraction of the algebra is given by,
\be \label{vir2bms}
L_k= \bL_k - \bL_{-k}, \ M_k = \e(\bL_k + \bL_{-k}).~~~\e\to 0
\ee
This is often known as an Ultrarelativistic (UR) contraction since the effective speed of light goes to zero in this construction. The resulting algebra is that of BMS$_3$, which is isomorphic to the Galilean Conformal Algebra (GCA) in two dimensions,
\begin{align}
\begin{split}
& [L_k, L_{k'}] = (k-k') L_{k+k'} + c_L\delta_{k+k',0} (k^3-k), \\&
 [L_k, M_{k'}] = (k-k')M_{k+k'} + c_M\delta_{k+k',0} (k^3-k), \\&
  [M_k, M_{k'}]=0. \label{bms}
\end{split}
\end{align}
Where $c_{L,M}$ are central charges to be determined.  At the level of coordinates and coupling constants, these correspond to singular scalings, viz.
\be \label{URlim}
\sigma \to \sigma, \ \tau \to \e\, \tau,  \ \e \to 0.
\ee
If one wants to relate the central charges, they also scale accordingly:
\be{}
c_L = c-\bar{c},~~c_M = \e (c+\bar{c}).
\ee
Starting from a CFT$_2$ action on flat spacetime and performing the above contraction leads one to the action,
\be \label{bmss}
S = \frac{1}{4\pi}\int~d\tau d\sigma (\partial_\tau \Phi)^2.
\ee
We note that only the temporal derivative of the field $\Phi(\sigma,\tau)$ survives under the contraction procedure. But there still survives the notion of space and time in the $(\sigma,\tau)$ coordinates, however they are not on equal footing as is the case of relativistic theories.
One could explicitly check that this action is invariant under the BMS transformations,
\be{}
\sigma \to f(\sigma), ~\tau \to f'(\sigma)\tau + g(\sigma).
\ee
Here $f,g$ are arbitrary functions and prime denotes a derivative w.r.t. $\sigma$. It is easy to see that these transformations are generated by the symmetry generators,
\bea{} 
&& L(f)=f'(\sigma)\tau\p_\tau+f(\sigma)\p_\sigma=\sum_k c_k e^{i\,k\,\sigma}(\p_\sigma+i\,k\,\tau\p_\tau)=-i\sum_k c_k L_k, \\
&& M(g)=g(\sigma)\p_\tau=\sum_k d_k e^{i\,k\,\sigma} \p_\tau=-i\sum_k d_k M_k\,, 
\eea
where $f=\sum c_k e^{i\,k\,\sigma},\ g=\sum d_k e^{i\,k\,\sigma}$ have been expanded in Fourier modes. The modes $L_n$ and $M_n$ generate the classical part of the BMS$_3$ algebra. The equations of motion for the scalar takes the form
\be \label{xeom}
\ddot{\Phi}=0.
\ee
Subject to periodic boundary conditions on a cylinder $\Phi(\tau,\sigma)=\Phi(\tau,\sigma+2\pi)$, the above EOM is solved by the following mode expansion:
\be \label{mode} 
\Phi(\sigma,\tau)= A_0+B_0\tau+\sum_{k}\frac{i}{k} \left(A_k-i\,k\,\tau B_k \right)e^{-i\,k\,\sigma}. 
\ee
Here $A, B$ are purely hermitian operators, and the conjugate momentum is given by
\be{}
\Pi = \frac{\p S}{\p \dot\Phi} = \frac{1}{2\pi}\dot\Phi.
\ee
and the canonical Poisson bracket which reads
\be{}
\{ \Pi(\sigma',\tau), \Phi(\sigma, \tau)\} = \delta(\sigma-\sigma'),
\ee
implies the following algebra for the oscillators,
\be\label{AB} 
\{A_k,A_{k'}\}_{P.B.}= \{B_k,B_{k'}\}_{P.B.}=0, \quad \{A_k,B_{k'}\}_{P.B.}= - 2\, i\, k\, \delta_{k+k',0}. 
\ee
These are clearly not usual CFT oscillators as is evident from the brackets, and they more look like Quantum mechanical oscillators $\{X,P\}$. However we can always go to a basis where these oscillators act as decoupled set of (anti) holomorphic oscillators \cite{Bagchi:2015nca},
\be \label{CC}
C_k = \frac{1}{2}({A}_k+B_{k}), \quad \C_k =\frac{1}{2}(-{A}_{-k}+B_{-k}).
\ee 
Now, the Poisson brackets take the canonical form: 
\be \label{4.22}
\{C_k, C_{k'} \} = -i\, k\, \delta_{k+k', 0} \ , \quad  \{ \C_{k},  \C_{k'} \} = -i\, k\, \delta_{k+k', 0} \ , \quad \{C_k, \C_{k'} \} = 0.
\ee
Starting from the generators
\be \label{lmab} 
L_k= \frac{1}{2} \sum_{k'} A_{- k'}B_{k'+k}\,\,\textrm{and} \quad M_{k}= \frac{1}{2} \sum_{k'} B_{-k'} B_{k'+k},
\ee
we can now write them in terms of the $C$ oscillators,
\bea{}
L_k&=&\frac{1}{2} \sum_{k'} \big[ C_{-k'} C_{k'+k} - \C_{-k'}\C_{k'-k} \big],\\
M_k&=&\frac{1}{2} \sum_{k'} \big[ C_{-k'}C_{k'+k} + \C_{-k'}\C_{k'-k} + 2\,C_{-k'}\C_{-k'-k} \big]. 
\eea 
These generators again span the BMS$_3$ algebra. However, one can spot that these generators are the same as the null string ones mentioned in \cite{Bagchi:2015nca}, however, with the spacetime indices stripped off. Many of our physical intuitions in subsequent sections will be borrowed from that of null strings, and we'll mention that in particular places. 
\subsection{Canonical quantization}
Let us try to understand the Hilbert space of the BMS invariant scalar theory.
As usual in quantized theory, all Poisson brackets go to Dirac brackets, and we can have the canonical commutation relations,
\be \label{c2}
[C_k,C_{k'}]=[\C_k,\C_{k'}]=k\, \delta_{k+k',0}.
\ee
And a CFT-like {\em{oscillator vacuum}} $\zc$ can be defined by the following
\be \label{c3}
C_k\zc=\C_k\zc=0\quad\forall\ k>0.
\ee
Here, one can clearly notice that $\zc$ is not a pure state anymore but an entangled state of these new Left and Right oscillator sectors:
\be{}
\zc = |0\rangle_R \otimes |0\rangle_L.
\ee
In this case, in terms of the $C$ oscillators, we can write down the relevant zero modes of the BMS generators:
\bea{}\label{c7}
L_0&=&\frac{1}{2} \sum_k \big[ C_{-k} C_{k} - \C_{-k}\C_{k} \big],\\
M_0&=&\frac{1}{2} \sum_k \big[ C_{-k} C_{k} + \C_{-k}\C_{k} + 2C_{-k}\C_{-k} \big]. 
\eea 
These can be thought of as analogues of angular momentum operator and Hamiltonian for usual relativistic CFT. However, note that $M_0$ here is seemingly not diagonalizable. This structure is central to defining the quantum nature of a BMS invariant theory. For details related to quantum structures and vacuum classifications of this theory, the reader is directed to \cite{Bagchi:2020fpr,Hao:2021urq}.

\subsection{Limiting perspective}
As we have emphasised earlier, BMS invariant theories can be discussed either from an intrinsic point of view, or equivalently by taking limits on their relativistic counterparts. 
To start along the second avenue, consider the relativistic free conformal scalar model on the cylinder, 
\be{} \label{rel_scalar_action}
S=\frac{1}{4\pi}\int d\sigma dt\Big((\partial_{t}\Phi)^2-(\partial_{\sigma}\Phi)^2\Big) ,\qquad(\sigma,\,t)\sim (\sigma+2\pi,\,t)
\ee
Under the UR limit (\ref{URlim}) together with the corresponding rescaling of the field,
\begin{equation}
t=\epsilon\, \tau,\quad
\Phi=\sqrt{\epsilon}\,\phi,\quad \epsilon\rightarrow0,  \label{model_UR_limit}
\end{equation}
the action \eqref{rel_scalar_action} becomes the BMS scalar action \eqref{bmss} on the cylinder $(\sigma,\tau)\sim (\sigma+2\pi,\tau)$, which we reproduce here,
\be{}
S=\frac{1}{4\pi}\int d\sigma d\tau\, (\partial_{\tau}\phi)^2\,.
\ee
The equation of motion of the relativistic scalar field (coming from (\ref{rel_scalar_action})) can be solved in terms of the mode expansion
\begin{equation}
\Phi(\s,t)=\phi_0+\pi_0\,t+\frac{i}{\sqrt{2}}\sum_{k\neq0}\frac{1}{k}(a_k\, e^{-i\,k\,(\sigma+t)} - \bar{a}_{-k}e^{-i\,k\,(\sigma-t)}).
\end{equation}
where $a_k^\dagger= a_{-k}$ etc., with the canonical commutation relations
 \begin{equation}
\label{combefore}[a_k,a_{k'}]=[\bar a_k,\bar a_{k'}]= k\,\delta_{k+k',0},\ \ [a_k,\bar a_{k'}]=0,\ \ [\phi_0,\pi_0]=i
\end{equation}
The CFT vacuum is defined by these oscillators
\be \label{c5}
a_k\za=\bar{a}_k\za=0\quad\forall\ k>0.
\ee
Comparing with the mode expansion of the BMS free scalar on the cylinder \eqref{mode}
we obtain the relation between modes before and after the UR limit
\bea{}
&A_k=\lim_{\epsilon\to 0}\frac{1}{\sqrt{\e}} \left( a_k - \bar{a}_{-k} \right),~ B_k=\lim_{\epsilon\to 0} {\sqrt{\e}} \left( a_k + \bar{a}_{-k} \right),~k\neq 0, \label{modeslimit}\\
&A_0 = \frac{\phi_0}{\sqrt{\epsilon}},~ B_0=-i\sqrt{\epsilon}\,\pi_0.
\eea
Following these limits and from \eqref{CC}, we can now see the relation between $C$ oscillators and the CFT oscillators in the limit read:
\begin{align}
C_{k}&=\frac{1}{2}\Big(\sqrt{\epsilon}+\frac{1}{\sqrt{\epsilon}}\Big)a_{k}+\frac{1}{2}\Big(\sqrt{\epsilon}-\frac{1}{\sqrt{\epsilon}}\Big)\bar{a}_{-k} \nonumber \\
\Tilde{C}_{k}&=\frac{1}{2}\Big(\sqrt{\epsilon}-\frac{1}{\sqrt{\epsilon}}\Big)a_{-k}+\frac{1}{2}\Big(\sqrt{\epsilon}+\frac{1}{\sqrt{\epsilon}}\Big)\bar{a}_{k}.
\label{infvel}
\end{align}
The general transformation between $C$ and $a$ oscillators turns out to be a Bogoliubov transformation since the canonical structure remains intact under the generic transformation:\footnote{Note that we could also have had
\be \label{c2}
C_k(\e) =\cosh \theta\, e^{-i\chi}\ a_k+\sinh \theta\, e^{i\chi} \ \bar{a}_{-k}, \quad \C_k(\e) =\sinh \theta\, e^{-i\chi} \ a_{-k}+\cosh \theta\, e^{i\chi} \ \bar{a}_{k}
\ee
which still would respect the canonical commutations as $\chi$ is just a pure phase. Squeezing operator in this case has to be changed accordingly. Since our $\e$ is considered purely real, we omit this extra phase factor.}
\be \label{c1}
C_k(\e) =\cosh \theta \ a_k-\sinh \theta \ \bar{a}_{-k}, \quad \C_k(\e) =-\sinh \theta \ a_{-k}+\cosh \theta \ \bar{a}_{k}. 
\ee
And quantum mechanically, the parameter changing from $\e = 1$ to $\e =0$ describes the contraction from a scalar CFT to a BMS scalar theory. At $\e = 0$ the oscillators explicitly belong to that of the BMS algebra, however the above relations hold even for $\e = 1$ where it goes back to the CFT oscillators \cite{Bagchi:2015nca}. Hence we can extrapolate these definitions for the whole range of validity for the parameter $\e$. Goes without saying, this is an approximation, but this helps us to understand the underlying structures better. The associated transformation can be generated using
\begin{align}
\begin{split} \label{CCC}
C_k = e^{-i G} a_{k} e^{iG},~~
\tilde{C}_k= e^{-i G} \bar{a}_{k} e^{iG}
\end{split}
\end{align}
Where the unitary transformation operator is a two-mode squeezing operator that can be written as,
\be\label{G}
G(\theta(\e)) = i \sum_{k=1}^{\infty} \frac{\theta}{k} \Big[a^{\dagger}_{k}\bar{a}^{\dagger}_{k} - a_k \bar{a}_k\Big],  \ee
 Remember that the new vacuum is defined by $C_k\zc=\C_k\zc=0\quad\forall\ k>0$, and this condition, using \eqref{c2} translates into:
\bea{bcond}
&&  (a_k -\tanh \theta\ \bar{a}_{-k})|0\rangle_{c}=0,\   k>0; \nonumber \\ 
&&  (\bar{a}_k-\tanh\theta\ a_{-k})|0\rangle_{c}=0.
\eea  

Now we are in a position to write down the mapping between the two vacua $\za$ and $|0\>_c$. This is given by the following two mode squeezed state:
 \begin{eqnarray} \label{zaa}
 |0\>_c = e^{-iG(\theta(\e))}\za
&=& \sqrt{\cosh\theta} \prod_{k=1}^{\infty}  \exp\left[\frac{\tanh\theta}{k} \ a^{\dagger}_{k} \bar{a}^{\dagger}_{k}\right]  |0\>_a.
 \end{eqnarray}
 Similarly, the inverse transformation to relate the two vacua reads,
\bea{bcond}
&& a_k|0\rangle_{a} = (C_k -\tanh \theta\ \tilde{C}_{-k})|0\rangle_{a}=0,\   k>0; \nonumber \\ 
&& \bar{a}_k|0\rangle_{a} = (\tilde{C}_k-\tanh\theta\ C_{-k})|0\rangle_{a}=0.
\eea  
Which can be thought to be generated by the inverse displacement operator:
\be \label{G2}
\bar{G}(\theta(\e)) = -i \sum_{k=1}^{\infty} \frac{\theta}{k} \Big[C^{\dagger}_{k}{\C}^{\dagger}_{k} - C_k {\C}_k\Big].
  \ee
  
With $C^\dagger_k = C_{-k}$ etc. Here we have $\tanh\theta=\frac{\e-1}{\e+1}$, which makes sure that (\ref{zaa}) is valid at $\e=1$.
The solution in this case is given as, 
\be \label{za2zc1}
|0\rangle_a = \frac{1}{\mathcal{N}} \prod_{k=1}^\infty \exp\left[- \frac{\tanh\theta}{k} C_k^\dagger\cdot\tilde{C}_k^\dagger\right]\zc. 
\ee
 Note that at the BMS (or the tensionless) point $\e=0$, the CFT vacuum turns out to be a special state w.r.t. the BMS oscillators:
\be \label{za2zc}
|0\rangle_a = \frac{1}{\mathcal{N'}} \prod_{k=1}^\infty \exp\left[- \frac{1}{k} C_k^\dagger\tilde{C}_k^\dagger\right]\zc. 
\ee
At the level of wavefunctions, the question is which way we want to evolve in $\e$. In a sense, this is a ``Thermal'' evolution and may be thought of as a Euclidean time evolution. More details on this can be found in \cite{Bagchi:2021ban}. 
\subsection{`Position space' representation of the vacuum }

In the present section, we will be computing circuit complexity for the state (\ref{zaa}). We  start by solving for the ``position-space" wavefunction. To do that, first let us define the following `position'  and `momentum' operators out of the $C$ oscillators,
\begin{align}\label{basis}
\begin{split}
& q_k =\frac{1}{\sqrt{2k}}\left(C^{\dagger}_{k}+C_k\right),\quad 
p_k = \frac{i}{\sqrt{2k}}\left(C^{\dagger}_{k}-C_k\right),\\&
\tilde{q}_k = \frac{i}{\sqrt{2k}}\left(\C^{\dagger}_k-\C_k\right), \quad
\tilde{p}_k = -\frac{1}{\sqrt{2k}}\left(\C^{\dagger}_k+\C_k \right),~~k>0. 
\end{split}
\end{align}
It is easy to check that they satisfy the canonical  commutation relations i.e, 
\be \label{comm}
[q_k,p_{k'}] = i\delta_{k,k'} = [\tilde{q}_k,\tilde{p}_{k'}].
\ee
 To do that,  we first write (\ref{bcond}) in terms of these position and momentum operators and then they give us the following first-order differential equations in position-space, which we can easily solve \footnote{We add a generic $k$ subscript to $\theta$. However, our Bogoliubov coefficients aren't directly mode dependent, so all $\theta_k$'s are the same.}:
\begin{align}
\begin{split} \label{solve}
&\Big(q_{k}+ i\, \tanh\theta_k\, \tilde q_k\Big)\psi_c(q_k,\tilde q_k)+\Big(\partial_{q_k}-i\, \tanh\theta_k\partial_{\tilde q_k}\Big)\psi_c(q_k,\tilde q_k)=0, \\&
\Big(i\, \tilde q_{k}-\, \tanh\theta_k\, q_k\Big)\psi_c(q_k,\tilde q_k)+\Big(i\, \partial_{\tilde q_k}+\tanh\theta_k\partial_{q_k}\Big)\psi_c(q_k,\tilde q_k)=0.
\end{split}
\end{align}
Note that, in the position space representation,  $$p_{k}=-i\,\frac{\partial}{\partial q_k}, \quad  \tilde p_{k}=-i\,\frac{\partial}{\partial \tilde q_k}.$$
Solving (\ref{solve}) we get the wavefunction, 
\begin{align} \label{neq2}
 \Psi_c(q_k,\tilde q_k)=\langle q_k;\tilde q_k | 0_c\rangle= \prod_{k=1}^{\infty}\frac{ e^{-A (q_k^2+\tilde q_k^2)- i\, B q_{k}\tilde q_k}}{\sqrt{\pi\, \cosh 2\theta_k}},
\end{align} 
where the constants are,
\bea && A=\frac{1}{2\, \cosh 2\theta_k}, \quad B= \tanh 2\theta_k. \label{eqqq}
\eea
Using the definition of $\theta_k$ i.e $ \tanh\theta_k=\frac{\e-1}{\e+1}$ we can rewrite these in the following manner, 
\bea
&&A=\frac{\e}{1+\e^2}, \quad B= \frac{\e^2-1}{\e^2+1}.
\eea
Further, we can introduce a new set of canonical variables to decouple the system into two sectors, 
\be \label{variablechange}
q^+_{k} =\frac{q_{k} +\tilde{q}_{k}}{\sqrt{2}}\,,~q^{-}_{k} =\frac{q_{k} -\tilde{q}_{k}}{\sqrt{2}}\,,~p^+_{k} =\frac{p_{k} +\tilde{p}_{k}}{\sqrt{2}}\,,~p^{-}_{k} =\frac{p_{k} -\tilde{p}_{k}}{\sqrt{2}}
\ee
Then the wavefunction mentioned in (\ref{neq2}) becomes,
\begin{align} \label{neq33}
\psi_c(q^{+}_k,q^{-}_k)=\prod_{k=1}^{\infty}\frac{ e^{-\frac{1}{2}(2\,A+i B)(q^{+}_k)^2-\frac{1}{2}(2\,A-i B)(q^{-}_k)^2}}{\sqrt{\pi\, \cosh 2\theta_k}}=\prod_{k=1}^{\infty}\psi^{+}_{k\,,c}(q^{+}_k)\psi^{-}_{k\,,c}(q^{-}_k),
\end{align}
In the next section, we will be using (\ref{neq33}) for the computation of circuit complexity. 
\section{Circuit complexity: From CFT to BMS} \label{sec2}
\subsection{Circuit complexity: A brief Introduction}
As mentioned before, our goal is to probe the transition from CFT to BMS using tools of quantum information. Particularly, we will focus on `\textit{circuit complexity}'.  We have already introduced this quantity in the introduction, but here let us give a brief technical review. We will mainly follow the approach pioneered by Nielsen, and his collaborators \cite{NL1,NL2,NL3}. For more details, interested readers are referred to \cite{Jefferson1}. Operationally, given a set of elementary gates, it quantifies the minimal number of operations needed to build a circuit which will take a suitable reference state  $| \psi_R \rangle$ as input and generate the desired target state $| \psi_T \rangle$ as an output. Formally, given a reference state and set of gates,  a quantum circuit starts at the reference state (at $s=0$) and terminates at a target state ($s=1$)
\begin{equation} \label{eqstate}
    |\psi_{T}(s=1)\rangle = U (s=1) |\psi_{T}(s=0)\rangle,
\end{equation}
Then 
where $U$ is the unitary operator that takes the reference state to the target state. It takes the following form, 
\begin{equation}
U(s)= {\overleftarrow{\mathcal{P}}} \exp[- i \int_0^{s} \hspace{-0.1in} ds' H(s') ] \ .
\end{equation}
The $s$ parametrizes a path in the space of the unitaries and given a set of elementary gates $M_I$, the  \textit{control} Hamiltonian ($H(s)$) can be written as
\begin{equation} \label{instgate}
    H(s)= Y^{I}(s) M_{I}\, .
\end{equation} 
The coefficients $Y^I(s)$  counts the number of times that a particular gate acts at a given value of $s.$ It can be easily shown that \cite{Jefferson1}, 
\begin{equation} \label{invert}
\frac{d U(s)}{ds} = -i\, Y^I(s) M_I U(s)\,.
\end{equation}
Then we define a cost functional $\mathcal{F} (U, \dot U)$ as follows:
\begin{equation}\label{length}
{\mathcal C}(U)= \int_0^1 \mathcal{F} (U, \dot U) ds\, .
\end{equation}
The dot defines the derivative w.r.t $s$. Minimizing this cost functional gives the optimal $Y^{I}$'s and hence it gives us the optimal circuit. There are different choices for the cost functional \cite{NL3,Jefferson1,Guo:2018kzl}. In this paper we will consider the following,
\begin{equation}
\mathcal{F}_2 (U, Y)  = \sqrt{\sum_I (Y^I)^2}\, .
\label{quadCost}
\end{equation}
 Here $\mathcal{F}_2 (U, Y)$ corresponds to standard distance measure over a  Riemannian geometry, here the one associated to the state space, on which \eqref{length} defines the length functional. \footnote{This is a natural choice for our study as we will  compare it with the Fubini-Study distance in a subsequent section.}\par

 For our case, a natural choice of the reference state $|\Psi_R\rangle$ is the CFT ground state which is a Gaussian state. So the reference wavefunction in $(q_k, \tilde q_k)$ basis takes the following form,
\begin{equation} \langle q_k;\tilde q_k |\psi\rangle_R=\psi_c(q_k,\tilde q_k)|_{\epsilon=1}= \prod_{k=1}^{\infty}\frac{ e^{-\frac{1}{2}(q_k^2+\tilde q_k^2)}}{\sqrt{\pi}}.
\end{equation}
 $\psi_c(q_k,\tilde q_k)$ is defined in (\ref{neq33}) and $\epsilon=1$ corresponds to the CFT ground state. Then for the target state  we choose the state mentioned in (\ref{neq33}) but for  $\epsilon \neq 1.$ In this way the circuit complexity will be function of the flow parameter $\epsilon$ and thereby will help us to probe the CFT to BMS flow. \par
\subsection{Behaviour of circuit complexity as a function of $\epsilon$}
Given the target and reference state we follow \cite{Camargo_2019, me1, Chapman:2018hou} to compute circuit complexity. Note that, both the target and reference state in our cases are Gaussian states. The Gaussian states are equivalently described by their corresponding `\textit{covariance matrix}'. The covariance matrix for each mode $k$ is defined in the following way,
\begin{align}
 \begin{split}
 G_k(\epsilon)=\langle \psi_c(q^{+}_k,q^{-}_k)|\varPsi_k \varPsi_k^{\dagger}|\psi_c(q^{+}_k,q^{-}_k)\rangle,
 \end{split}
    \end{align}
    where, $$\varPsi^{T}=\Big\{q_{k}^{+},p_{k}^{+}, q_{k}^{-}, p_{k}^{-}\Big\}.$$
    For our case we will have the following two covariance matrices for reference ($\epsilon=1$) and target state ($\epsilon \neq 1$) wavefunctions, 
    \begin{align}
    \begin{split}
        G_k^{s=0}(\epsilon=1)=\begin{pmatrix} 1 &0&0&0\\ 0&1&0&0\\0&0&1&0\\0&0&0&1 \end{pmatrix},\quad G_k^{s=1}(\epsilon)=\begin{pmatrix} \frac{1}{2 A} &-\frac{B}{2 A}&0&0\\ -\frac{B}{2 A}&\frac{4 A^2+B^2}{2 A}&0&0\\0&0&\frac{1}{2 A}&\frac{B}{2 A}\\0&0&\frac{B}{2A}&\frac{4 A^2+B^2}{2A} \end{pmatrix}\,.
        \end{split}
    \end{align}
    We note that, from (\ref{eqqq}) $4 A^2+B^2=1.$ We can compute the circuit complexity in terms of these covariance matrices \cite{Camargo_2019,me1}. We want to construct the optimal circuit such that, 
    \begin{align}
        \begin{split}
            G_k^{s=1}= U(s=1)\cdot G_k^{s=0}\cdot U(s=1)^{T}.
        \end{split}
    \end{align}
   Note that, these covariance matrices are of block diagonal form. Each of the blocks are an element of $SU(1,1)$ group. So we can take the generators $M_I$'s as generators  $SU(1,1) \times SU(1,1)$. For details interested readers are referred to \cite{me1}. Finally the complexity per mode $k$ takes the following form  due to the structure of the covariance matrix,\ \cite{me1} \footnote{Note that, the total complexity will be $\mathcal{C}=\sqrt{\sum_{k=1}^{N}\Big(\textrm{arccosh}\Big(\frac{1}{2 A}\Big)\Big)^2}.$ As the argument of $\textrm{arccosh}$ is independent of $k,$ we will get, $\mathcal{C}=\frac{1}{\sqrt{2}}|\textrm{arccosh}\,\,\frac{1}{2 A}\Big|\, \sqrt{V}$, where $V$ is the momentum space volume i.e $V=\sum_{k=1}^{N}.$ This overall factor of $V$ do not affect our conclusions, hence we focus on the complexity per volume to avoid unnecessary clutter.}
    \begin{align}
        \mathcal{C}_k= \frac{1}{\sqrt{2}}\,\Big|\textrm{arccosh}\Big(\frac{1+4 A^2+B^2}{4 A}\Big)\Big|=\frac{1}{\sqrt{2}}\,\Big|\textrm{arccosh}\,\,\frac{1}{2 A}\Big|=\frac{1}{\sqrt{2}}\,\Big|\textrm{arccosh}\,\,\frac{1+\epsilon^2}{2\,\epsilon}\Big|\,.
        \end{align}

\begin{figure}[ht!]
\begin{center}
\scalebox{0.6}{\includegraphics{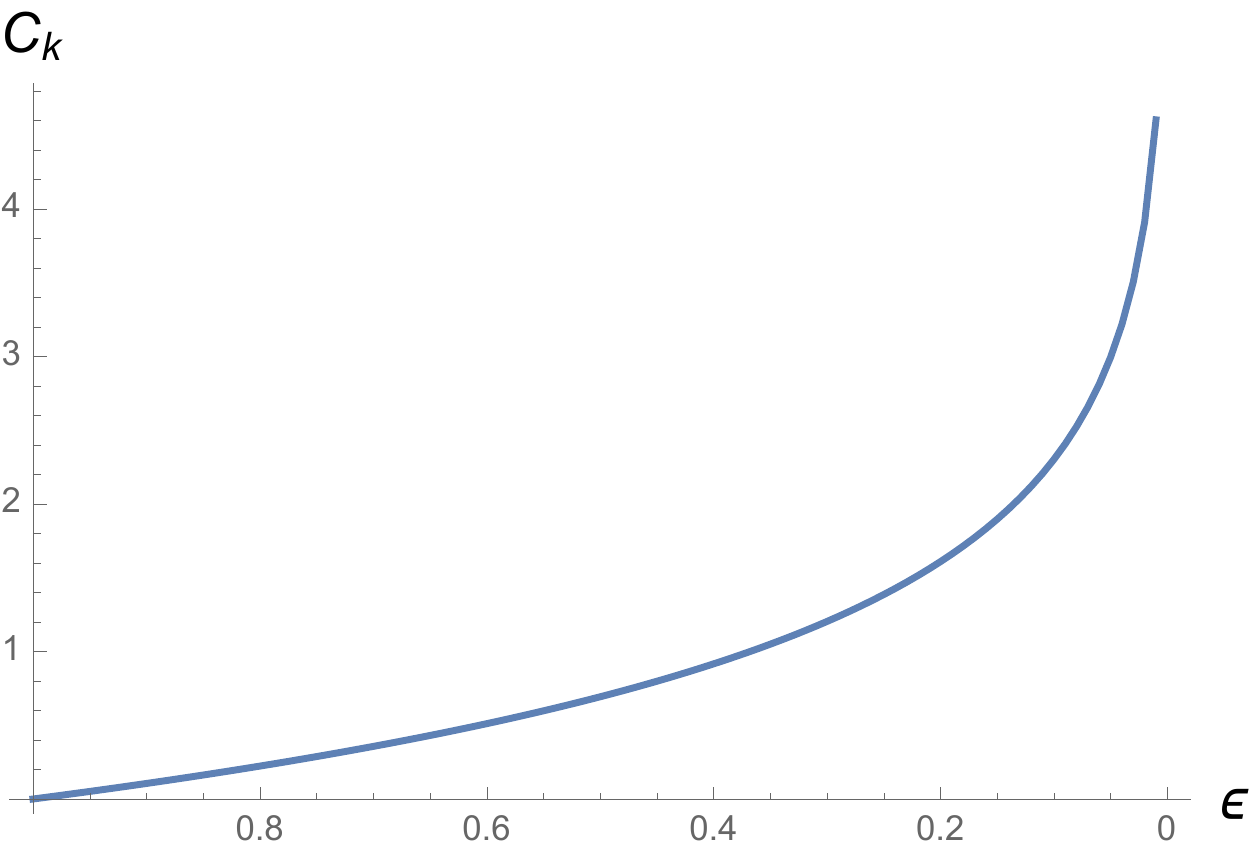}}
\end{center}
\caption{Complexity as function of flow parameter $\epsilon.$ $\epsilon=1$ and $\epsilon=0$ corresponds to the CFT and BMS point respectively. It clearly diverges at the point $\epsilon=0.$ We have rescaled the complexity by a factor of $\sqrt{2}.$}
\label{fig:regular}
\end{figure}

It is evident that $C_{k}$ is a monotonically \textit{increasing} function of the parameter $\epsilon.$ It starts from zero  at $\epsilon=1$, i.e. at the CFT ground state and diverges at $\epsilon=0$ i.e. at the BMS vacuum. This is illustrated in the Figure.~\ref{fig:regular}.

Some comments are in order after this result. The divergence in circuit complexity indicates that the target state may not be reachable from the reference state via a combination of unitary operations. But this can also be interpreted as nonanalyticity corresponding to some critical points \cite{Liu:2019aji} and signals the presence of a quantum phase transition. Looking at the system at hand, it makes perfect sense to assume there is a phase transition in going from CFT to BMS at the very extreme point, where a notion of ultralocality sets in. For the tensionless string case, this phase transition was interpreted as a Bose-Einstein like condensation that gives rise to open strings degrees of freedom from closed strings \cite{Bagchi:2019cay}, as the target state is essentially a boundary state along with all spacetime directions. We can safely assume a related interpretation for our case as well. However, the actual physical perspective may be different here.

\section{Information Geometry} \label{sec3}
\subsection{Fubini-Study metric}
\begin{figure}[t!]
\begin{center}
\scalebox{0.6}{\includegraphics{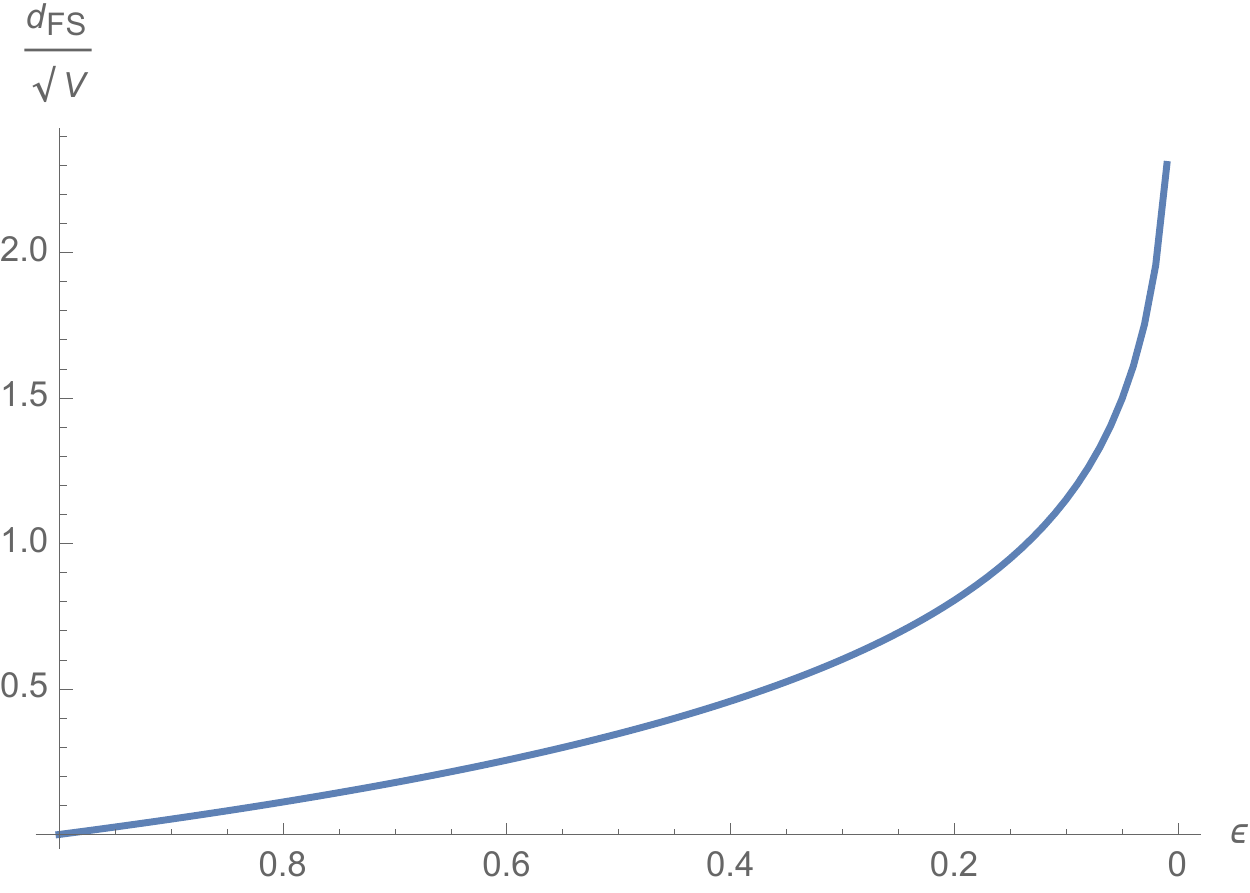}}
\end{center}
\caption{Fubini-Study distance as function of flow parameter $\epsilon.$ $\epsilon=1$ and $\epsilon=0$ corresponds to the CFT and BMS point respectively.}
\label{fig:regular1}
\end{figure}

To get further insight into the diverging complexity at $\epsilon=0,$ as we have uncovered in the last section, we will first try to associate a Riemannian structure to the space of wavefunctions (\ref{zaa}). We identify the coherent state we have been working with a  point on a group manifold, and the complexity for the target state is defined as the geodesic distance between the state and the reference one on the group manifold. 
As noted before the state mentioned in (\ref{zaa}) is a $SU(1,1)$ coherent state. So we start with a generic state of the form, 
\be \label{coherentstate}
|\psi\rangle=\mathcal{N}\prod_{k=1}^{N} e^{z_k\, K_{+}}|0,0\rangle.
\ee
Here $K_+$ is a $SU(1,1)$ generator, with the set of generators given by combination of oscillators (without the $k$ subscripts) :
\be\label{su11}
K_{+}=\beta^\dagger \tilde{\beta}^\dagger,~K_{-} = \beta\tilde{\beta},~ K_{z} = \frac{1}{2}\left(\beta^\dagger\beta+ \tilde{\beta}\tilde{\beta}^\dagger   \right),
\ee
which generates the familiar algebra:
\be{}
[K_+, K_-]=-2K_z,~[K_z, K_\pm]=\pm K_\pm.
\ee
Then the state associated to \eqref{coherentstate} can be given a Riemannian structure \cite{Provost:1980nc}. The infinitesimal distance in this state space, also know as \textit{`Fubini-Study'} metric, can be written as
\be
ds^2= g_{ij} dx^{i}dx^{j},
\ee
where the metric tensor is given by, 
\be
g_{ij}=\langle \partial_i\psi|\partial_j\psi\rangle-\langle\partial_i \psi|\psi\rangle\langle\psi|\partial_j\psi\rangle.
\ee
Finally we get \cite{Chapman:2017rqy,Ali:2018aon}, 
\be
ds^2=\sum_k \frac{ |dz_k|^2}{(1-|z_k|^2)^2}.
\ee
Further we can paramterize the complex function $z_k$ in \eqref{coherentstate} as $$z=|z_k|e^{i\,\phi_k},$$ where we take $|z_k|=\tanh(\bar{\theta}_k/2).$ Then we get, 
\be \label{metric}
ds^2=\sum_{k=1}^{N} \frac{1}{4}\Big(d\bar{\theta}_k^2+\sinh(\bar{\theta}_k)^2d\phi_k^2\Big).
\ee
Then the geodesic distance between two point $(\bar{\theta}_{1\,,k},\phi_{1\,,k})$ and $(\bar{\theta}_{2\,,k},\phi_{2\,,k})$ is given by, 
\be
d_{FS}=\frac{1}{2}\sqrt{\sum_{k=1}^{N}\Big(\textrm{arccosh}\Big[\cosh(\bar{\theta}_{1,k})\cosh(\bar{\theta}_{2,k})-\sinh(\bar{\theta}_{1,k})\sinh(\bar{\theta}_{2,k})\cos(\phi_{1,k}-\phi_{2,k})\Big]\Big)^2}.
\ee
From (\ref{zaa}) it is evident that $z_k=\frac{(\epsilon-1)}{ (\epsilon+1)}$. Also, we can clearly see that for CFT ($\epsilon=1$) $z_k=0$ as $\bar \theta_k=0$ and for BMS ($\epsilon=0$) $z_k=-1.$ Hence the length of the geodesic connecting the following two points \footnote{Note that, although the phases $\phi_k$ are zero for the in initial and final state, the shortest geodesic connecting them could pass through states with non-vanishing phase. So for computations pertaining to an intermediate state, we should keep track of the phase factor.}, $$(\bar{\theta}_{1,k}=0,\,\, \phi_{1,k}=0)\,,\quad (\bar{\theta}_{2,k}=2\,\textrm{arctanh}\Big(\frac{(\epsilon-1)}{\, (\epsilon +1)}\Big),\,\, \phi_{2,k}=0)\,\, [\epsilon < 1],$$\\
turns out to be,
\be
d_{FS}=\frac{1}{2}\sqrt{\sum_{k=1}^{N}\bar{\theta}_{2,k}^2}=\sqrt{\sum_{k=1}^{N}\textrm{arctanh}\Big(\frac{(\epsilon-1)}{\, (\epsilon +1)}\Big)^2}=\textrm{arctanh}\Big(\frac{(\epsilon-1)}{\, (\epsilon +1)}\Big)\sqrt{V}.
\ee
Here $V=\sqrt{\sum_{k=1}^{N}}$ denotes the phase-space volume. It is easy to check that it is a monotonically increasing quantity and diverges at $\epsilon=0.$ This is shown in the Figure.~\ref{fig:regular1}. Also, note that, for $\epsilon=0,$ $z_k=1$  and from (\ref{metric}) it is evident that the information metric becomes degenerate as we approach the BMS point, i.e. the geodesic never reaches the BMS point staying on the same coordinate chart.


So we could again see that the complexity diverges at the special point of $\e = 0$ as before, showing similar qualitative behaviour. Although the geometric notion associated with this intriguing observation is still unclear, one could recall that $|z_k| =1$ corresponds to a degenerate point on the projective space hyperbola on the Fubini Study metric. This also corresponds to the phase transition point in the physical space, especially with the ones associated with ground state degeneracies. As seen in the literature \cite{Bagchi:2019cay} this particular point with $\e=0$ has been interpreted as an infinitely degenerate vacuum with all excitations in the tensile string condensing into just one state. This could be the real reason behind this diverging geodesic distance. However, a true CFT notion is beyond the scope of this work.

\subsection{Berry Curvature}
Furthermore, we can compute the Berry curvature \cite{Berry:1984jv} associated with these kinds of $SU(1,1)$ coherent state (\ref{coherentstate}), transforming finally into a boundary state in this process. The Berry curvature is a measure to quantify a path in a group representation that connects our initial and final points. The components of Berry connections are defined by, 
\be \label{Berrc}
A_i=\langle \psi| \partial_{i} |\psi\rangle\,.
\ee
For our case, we get the following components for each $k$ mode \cite{Nogueira:2021ngh}, 
\be
A_{z_k}=\frac{\bar {z}_k}{2\,(1-|z_k|^2)},\quad A_{\bar z_k}=-\frac{{z}_k}{2\,(1-|z_k|^2)}\,.
\ee
 The bar on $z_k$ denotes the complex conjugate for generic parametrisations. Then we can define a two-form, the Berry curvature, for each mode $k$ as follows, 
 \be
 F= d A
 \ee
where $d$ is the exterior derivative and the one-form $A$ is Berry connection components which are defined in (\ref{Berrc}). For our case, we will have, 
\be
F=\frac{i}{2}\sinh(\bar \theta_k)\, d\theta_k\wedge d\phi_k\,.
\ee
Here we have used the fact that $z=\tanh(\bar{\theta}_k/2) e^{i\,\phi_k}.$ For the state (\ref{zaa}), using $\bar{\theta}_{k}=2\,\textrm{arctanh}\Big(\frac{(\epsilon-1)}{\, (\epsilon +1)}\Big)$ as before we get,
\be
F_{\theta\phi}=\frac{i}{2}\sinh \Big[2\,  \textrm{arctanh}\Big(\frac{(\epsilon-1)}{\, (\epsilon +1)}\Big)\Big]=\frac{i}{4}\Big(\frac{\epsilon ^2-1}{\epsilon }\Big).
\ee
It is easy to see that the Berry curvature diverges at $\epsilon=0$, i.e. at the BMS point. So the behaviour of both the complexity and the Berry curvature is the same at the critical point $\e =0$.

\section{Hamiltonian Evolution} \label{sec4}
For the last couple of sections, we have been focusing on the evolution of our system only in $\e$,  presumably at an initial timeslice. Our discussion clearly shows that the quantities we are interested in could be ill-defined at the pure BMS point $\e=0$. In this sense, $\e$ acts as a cut-off in the system. However, the story could be different at a finite timeslice, which we will deliberate on in this section.

\subsection{Diagonalisation}

Let us now consider the temporal dynamics associated with our system's Hamiltonian that also changes with the parameter $\e$. One can remember that for relativistic $2d$ CFT, the Hamiltonian and Angular momentum operators were tentatively given by combinations of Virasoro zero modes,
\be{}
H_M=\mathcal{L}_0+\bL_0,~~~J_M=\mathcal{L}_0-\bL_0.
\ee
Upon quantization, the operators $\mathcal{L}_0$, $\bL_0$ are given in terms of CFT oscillators \eqref{c5} by 
\be{} \mathcal{L}_0=\frac{1}{2}\sum_k a_{-k}a_{k},~~
\bL_0=\frac{1}{2}\sum_k \bar{a}_{-k}\bar{a}_{k} \quad, \ee
Now remember that under contraction, the oscillators change with a Bogoliubov transformation,
which are the inverse transforms of (\ref{infvel}),
\bea{}
a_k&=\Omega_+ C_k - \Omega_-\tilde{C}_{-k}\nonumber \\
\bar{a}_k&=\Omega_+\tilde{C}_{k}-\Omega_- C_{-k}.
 \eea
Using the above, we can see that action of the operator $(\mathcal{L}_0^{(\e)}-\bL_0^{(\e)})$ on a state remains invariant throughout the $\e$ evolution since $\Omega_+^2-\Omega_-^2=1$ by definition of Bogoliubov transformations. But the other combination reads,
\be \label{hami2}
\mathcal{L}_0^{(\e)}+\bL_0^{(\e)}=\sum_k\left[ ( \Omega_+^2+ \Omega_-^2)( C_{k}^\dagger C_{k}+\tilde{C}_{k}^\dagger \tilde{C}_{k} )-4\,\Omega_+ \Omega_- C_k\tilde{C}_k   \right].
\ee
Note that this is the Hamiltonian that appeared in \cite{Bagchi:2021ban} in the null string theory context. So the action of $\bL_0^{(e)}+\bL_0^{(e)}$ combination does not remain invariant as we move to more and more in $\e$, and an extra ``perturbation'' term generates a deformation \footnote{See \cite{Bagchi:2022nvj} for some physical insight into the nature of this term as a current-current deformation to the CFT.}. This is the extra seemingly non-diagonal term in the above equation.\par 
A possible way out of the problem is to consider the Hamiltonian arbitrary near the null surface, where the BMS symmetry arises. Here, we can write an appropriately scaled and finite perturbative normal ordered Hamiltonian near $\e\to 0$ (but not exactly) with next to leading correction in $\e$, \footnote{ There is an implicit $\e$ multiplying the whole Hamiltonian to make it finite, i.e. $H\to\e H$.}
\be \label{cH} 
H_{\e} = \sum_{k=0}^{\infty} \left[ (1+\e^2)( C_{-k}C_{k}+\tilde{C}_{-k} \tilde{C}_{k} )+ (1-\e^2)( C_{-k}\tilde{C}_{-k}+C_{k} \tilde{C}_{k})\right].
\ee
 
We can see that this Hamiltonian consists of two normal ordered number operators and two non-diagonal parts. As we go to $\e= 0$, we get back the exact BMS answer for $M_0$ in (\ref{c7}). However, as we saw for the bogoliubov transformations, this definition can also be extrapolated to the CFT point $\e =1$, where only the two number operators remain (with the identification $a=C$ etc). Note that, with the definitions of the basis we use in \eqref{basis}, the perturbation term in the Hamiltonian can be written as
\begin{align}
C^{\dagger}_{k}\tilde{C}^{\dagger}_{k} + C_k \tilde{C}_k = -k\left(q_k\tilde{p}_k + p_k\tilde{q}_k  \right).
\end{align}
We can further notice that these commutation relations \eqref{comm} are invariant upto a discrete transformation,
\be\label{flip}
\tilde{p}_k \to -\tilde{q}_k,~~\tilde{q}_k\to \tilde{p}_k,
\ee
or similarly for non-tilde variables, which basically gives another set of basis oscillators for our wavefunction, where the tilde and non-tilde set of $(q,p)$s in \eqref{basis}  are treated in the same footing. These transformations can also be achieved by a `flipping' map of $\C$ oscillators $\C_k \to \C'_k= i\C_{-k}$. Evidently, the perturbation term in the Hamiltonian also changes under this map:
\be{}
k\left(q_k\tilde{q}_k - p_k\tilde{p}_k \right) = i\Big(C^{\dagger}_{k}{\C}^{\dagger}_{k} - C_k {\C}_k\Big),
\ee
which is just the displacement operator, and usually appears in the Hamiltonian for a Thermo Field Double (TFD). \par 
Next we time-evolve (\ref{neq2}) with the Hamiltonian (\ref{cH}) written in the position-momentum basis.   After neglecting a constant additive term (\ref{cH}) becomes, 
\begin{align} \label{neq3}
H_{\epsilon}=\sum_{k=0}^{\infty}\Big[\frac{k}{2}(1+\epsilon^2)(q_k^2+p_k^2+\tilde q_k^2+\tilde p_k^2)-k (1-\epsilon^2)\left(q_k\tilde{p}_k + p_k\tilde{q}_k  \right)\Big].
\end{align}
Furthermore we can diagonalize (\ref{neq3}) by using the  transformations as below:
\be \label{variablechangev2}
q^+_{k} =\frac{\tilde{q}_{k} -{p}_{k}}{\sqrt{2}}\,,~q^{-}_{k} =\frac{\tilde{q}_{k} +{p}_{k}}{\sqrt{2}}\,,~p^+_{k} =\frac{q_{k} +\tilde{p}_{k}}{\sqrt{2}}\,,~p^{-}_{k} =-\frac{q_{k} -\tilde{p}_{k}}{\sqrt{2}},
\ee
which are related to our earlier expression in \eqref{variablechange} via the identifications in \eqref{flip}. Then in terms of these new $(\pm)$ variables  the hamiltonian becomes: 
\begin{equation} \label{diagonalized}
    H_{\epsilon} = \sum_{k=0}^{\infty}k\left[\e^2(p^{+^2}_{k}+q^{-^2}_{k})+p^{-^2}_{k}+q^{+^2}_{k}  \right]
\end{equation}
One can also see here that imposing \eqref{flip} into the Hamiltonian and diagonalizing it gives rise to the same structure as above. As one can see from (\ref{diagonalized}), in the strict $\e = 0$ limit, both oscillators ``freeze out'', i.e. there are no dynamics at all. One may be tempted to call this a true Carrollian situation, where lighcones close down, and there is no movement in space at all. We should moreover note from (\ref{diagonalized})that there are two different sets of eigenvalues of this Hamiltonian. One set scales as $\epsilon^2$ and vanishes in the limit $\epsilon=0$, i.e. at the BMS point, and the other set scales as $\frac{1}{\epsilon^2}$ which survives at the BMS point, effectively leading to one remaining set of oscillators. 
\subsection{Out-of-time-ordered correlators}
Let us now actually focus on a particular observable diagnostic of quantum chaos, namely Out-of-time-ordered correlators (OTOCs) for this system, armed with our diagonal Hamiltonian. This will also require us to talk about how operators time evolve in this system as $\e$ changes. In general OTOCs in a quantum system are defined as $ C_T(t) = -\langle[ W(t),V(0)]^2\rangle$, where $W(t)$ and $V(0)$ are some generic operators in Heisenberg representation at time `$t$' and some initial time.  Let's then start with the diagonlised Hamiltonian mentioned in (\ref{diagonalized}). Time evolution with this at $\e=0$ is tricky as there is no apparent dynamics, hence we need to calculate our OTOCs at finite (but small) $\e$, at finite time, and take a suitable limit.\par 
We choose the position and momentum operators in the $\pm$ basis (\ref{variablechange}) as of interest. Under time evolution, the operators change as following \cite{Hashimoto:2017oit}:
\bea{}
 q^{\pm}_{k}(t) &=& \cos(2\,k\,\epsilon\, t)~q^{\pm}_{k}(0) + \e^{\pm 1} \ \sin(2\,k\,\epsilon\, t)~p^{\pm}_{k}(0)\\
    p^{\pm}_{k}(t) &=& \cos(2\,k\,\epsilon\, t)~p^{\pm}_{k}(0) - \e^{\mp 1} \ \sin(2\,k\,\epsilon\, t)~q^{\pm}_{k}(0)
  \eea
The OTOCs in this case is the given by:
\begin{eqnarray}\label{otoc}
[q^{\pm}_{k}(t),q^{\pm}_{k}(0)] &=& i\epsilon^{\pm 1} \sin(2\,k\,\epsilon\, t)\nonumber\\
\left[p^{\pm}_{k}(t),p^{\pm}_{k}(0)\right]&=&  i\epsilon^{\mp 1} \sin(2\,k\,\epsilon\, t)\nonumber\\
\left[q^{\pm}_{k}(t),p^{\pm}_{k}(0)\right]&=& i \cos(2\,k\,\epsilon\, t)
\end{eqnarray}
One can observe that while at finite values of $\e$ the OTOCs scale sinusoidally, in the strict limit of $\e\to 0$ they either go to zero or scale \textit{polynomially} with time ($k^2 t^2$ to be exact), signalling the freeze-out we just discussed \footnote{Following \cite{Gharibyan:2018fax,Bhattacharyya:2020art}, one can also calculate the entire Lyapunov spectrum. One first constructs the matrix $ L= M^{\dagger} M,$ where $M_{ij}=i[z_{i}(t), z_{j}(0)]$ with $i,j=1,2,3,4$  and $z =\{q^{+}, q^{-}, p^{+},p^{-}\}$ for each value of the mode $k$. Then the eigenvalues of $L$ give the information about the entire Lyapunov spectrum. For a chaotic system, these eigenvalues typically behave as the exponential of $t$, and the exponents give the quantum Lyapunov spectrum. For our case, we can easily check that for $\epsilon=0,$ the eigenvalues of $L$ are polynomials of $t$, indicating the absence of chaotic behaviour.}. Note also that while the bracket $[q(t),p(0)]$ gives the canonical commutation relation at $t=0$, the same behaviour comes back at $\e=0$ too. This is an intriguing dynamical behaviour, as the Lyapunovian exponential behaviour gives way to this polynomial growth. However, this phenomenon and its consequences need to be understood in a better physical way which we leave for future work. 
\subsection{Krylov complexity}
In this section, to get further insight into the dynamics of the system, we sketch the idea of the complexity of Hamiltonian evolution. In this context, a natural notion of complexity which has been investigated in recent times in various contexts is \textit{Krylov Complexity} \cite{Parker:2018yvk}. In recent times, \textit{operator growth} has played an important role in the context many-body system \cite{vonKeyserlingk:2017dyr,Nahum:2017yvy,Khemani:2017nda,Chan:2017kzq}. An operator grows under the Liouvillian superoperator, and Krylov Complexity captures the notion of the spread of the operators. 
\par  To proceed, let us think of our coherent state in the context of operator evolution under $SU(1,1)\approx SL(2,R)$, where the states are written as:
\be
|\psi\rangle  = D(\xi)|0\rangle,~~~D(\xi)= \prod_{k}e^{\xi_k L_{-1}-\bar{\xi_k}L_{1}}.
\ee
For our case, $\xi=\bar{\xi}$ is a constant real parameter, and the two-mode squeezed state representation of $L$ operators are analogous to \eqref{su11} i.e. we can identify $L_{\mp 1} = K_{\pm},~L_0= K_z$.  In the case where $\xi=\bar{\xi}$ is complex and proportional to time, this signifies unitary evolution with the Hamiltonian. However, in this case, our total hamiltonian \eqref{cH} is given by 
\be\label{ham}
H = \gamma_1 (L_1+L_{-1})+\gamma_2 L_0,
\ee
where the coefficients are real, $\gamma_2 =2 (1+\e^2)$ and $\gamma_1=(1-\e^2)$. This is a generic $SL(2,R)$ Hamiltonian, albeit we do not have a unity component as it would just contribute an overall phase. It can always be restored by suitably normal ordering the Hamiltonian. Time evolution under this can be thought of as producing generalized time-dependent coherent states. Notice again that at $\e =1$, i.e. at the CFT point, $\gamma_1=0$, and there is no generic displacement operator at work.

This being said, we can consider this evolution as the time-dependent evolution of the thermofield double state. Here two copies of the Virasoro CFT were disjoint at first, but they start talking to each other once $\e<1$ and produce a maximally entangled (boundary) state at $\e = 0$. It has been argued \cite{Bagchi:2015nca} that the interpolating vacuum during $\e$ evolution, i.e. $|0\>_c$ \eqref{zaa} signifies a thermal phase of the CFT. This was further corroborated in recent works \cite{Bagchi:2020ats, Bagchi:2021ban} concerning null strings where this vacuum was interpreted as the vacuum for an analogue of the worldsheet Unruh effect, driven by the Bogoliubov transformations \eqref{infvel} near the extreme. Here the parameter $\e$ sets the scale for inverse acceleration and hence the same for inverse temperature.

If this interpretation withstands, the generic thermal evolving state at an initial time can be written as:
\be
|0\>_c= \mathcal{N}\sum_{k}~e^{-\beta \omega_k/2}|k\>\otimes |\tilde{k}\>,~~\omega_k = k+\frac{1}{2}.
\ee
Here $\beta$ is as usual the inverse temperature.
It can then be shown using the Lanczos algorithm \cite{Caputa:2021sib} by assuming a particular representation of the state space that unitary time evolution of the above state under \eqref{ham} requires the strength parameters to have a form,
\be
\gamma_1 = \frac{\omega}{2\sinh{\frac{\beta\omega}{2}}},
~~\gamma_2= \frac{\omega}{\tanh{\frac{\beta\omega}{2}}}.
\ee
Note here, since the frequencies are generally $k$ dependent, the $\gamma$s should also be $k$ dependent. However, the explicit one-parameter form of the coefficients prohibits that, and we take $\omega_k = \omega$, i.e. we concentrate on a single mode, without any $k$ dependence. Now the time evolved TFD state is analogous to $e^{iHt}|0\>_c$, which we can compute using the Baker–Campbell–Hausdorff formula, and that is the target state we are looking for. The characteristic oscillation frequency for our case is then set as:
\be
\frac{\omega^2}{4} = -\gamma_1^2+ \frac{\gamma_2^2}{4}=4\e^2.
\ee
Now one can see that there are clearly three dynamical regimes for our system. For $\gamma_2 > 2\gamma_1$, the frequency is real, while for  $\gamma_2 < 2\gamma_1$, the frequency is imaginary. One can think of these two regimes respectively corresponding to the standard and inverted harmonic oscillators. The transition point between these two regimes  $\gamma_2 = 2\gamma_1$ is intriguing for us as $\e=0$ at this point. This makes sense as the frequency becomes zero at this point, reinforcing our comments on the freeze-out of dynamics at the onset of Carrollian physics. 

Following the discussion in \cite{Caputa:2021sib}, we can find that the Krylov basis is the standard two-oscillator Fock space and Krylov complexity is proportional to the average particle number in the time evolved state: 
\be
C(t)  \propto\frac{1}{\left(1-\frac{\gamma_2^2}{4\gamma_1^2}\right)}\sinh^2\left(\sqrt{\gamma_1^2- \frac{\gamma_2^2}{4}}~t  \right)= { \gamma_1^2\Big(\frac{\sin(2\e\, t)}{2\e}\Big)^2}
\ee
which grows exponentially when  $\gamma_2 < 2\gamma_1$ and has the usual sinusoidal behaviour when  $\gamma_2 > 2\gamma_1$. For our system, notice that when $0<\e\leq 1$
this quantity will always vary as a sinusoid. Explicitly at the BMS point $\e = 0$, we have a vanishing frequency and hence the Complexity varies quadratically with time, i.e.
\be
\lim_{\gamma_1\to 2\gamma_2} C(t) \sim \gamma_1^2 t^2,
\ee
which has a similar scaling as our OTOCs in \eqref{otoc} for the system at the BMS point. This result remains finite even at $\e = 0$ when we take the limit carefully.

Before ending this discussion, let us also notice an intriguing fact about the physical significance of the parameter space region $\gamma_2 < 2\gamma_1$ for our system. This explicitly points to the situation where $\e \to i\e$, i.e. a complex contraction of the conformal algebra \footnote{One may also be tempted to interpret the scaling $t \to i\e \tau $ as equivalent to contracting an \textit{euclidean} theory.}. In this regime, one could have an exponentially growing behaviour of the complexity, commensurate with an unstable phase of the oscillator, and consequently, a Lyapunov exponent can be read off \cite{Ali:2019zcj,Bhattacharyya:2020art,Bhattacharyya:2021cwf, Balasubramanian:2022tpr} \footnote{Several other works have investigated whether complexity can detect the scrambling time and Lyapunov exponent, e.g. \cite{haqueReducedDensityMatrix,Ryuchaos,Magan2018-ua,Qu:2021ius,Balasubramanian:2019wgd,Yang:2019iav}. This list is by no means exhaustive, and a thorough look at the reference and citations of these papers is recommended.}. 
\section{Discussions and Conclusions} \label{sec5}
In this paper, we discussed information-theoretic probes for the transition from a CFT$_2$ to a BMS$_3$ invariant scalar field theory. It is well known that these two theories are related via an Inönu-Wigner contraction, which we explicitly used to construct quantum states that flow from one theory to another. It is also well documented that the endpoint in this path, the BMS invariant theory, presents singularities and degeneracies associated with the Carrollian manifold it inherently lives on. We took our reference state as the Gaussian ground state of the CFT and the target state as the entangled ground state of BMS, which turned out to be related to each other via a squeezing operation. Since at the exact BMS point, the state evolves into a boundary state, the underlying physics is expected to change drastically, and our computations bear witness to this fact. We explicitly showed that the complexity diverges at this critical point, signalling a quantum phase transition into a unitarily inequivalent theory. We proceeded to show that these two states are connected via an infinite length geodesic on the state manifold, which proves what was said before. The same behaviour was reproduced when we extracted the Berry Curvature associated with this process, effectively indicating the cut-off nature of the contraction parameter. 

To understand this transition better, we then quantified the time evolution under the total Hamiltonian of the system, which continuously varies with the contraction parameter. Time-dependent markers of quantum chaos turn out to be much better controlled when a careful $\e \to 0$ limit is taken on them. It turns out that dialling the contraction parameter from CFT to BMS changes the OTOCs of the system from oscillatory to polynomial behaviours. We also looked at the operator complexity associated with this transition and found it to scale polynomially with time as well. It was very important to note that when $0<\e\leq 1$  the complexity varies sinusoidally, while at the transition point $\e = 0$ it reduces to scale with $t^2$, signifying two completely different phase structures associated with these realms. All of these results point to the apparent absence of chaotic behaviour in this transition.

It would be nice to understand the origin of the polynomial behaviour for the time-dependent quantities we talked about in this work. It is intriguing to see that such systems have been discussed in the literature (see, e.g. \cite{Kuwahara:2020chn}), and point out a regime where fast scrambling may not be present for the system. However, a connection is merely speculation at this point, and a concrete mathematical link has to be established rigorously.

One could also go ahead and ask whether similar physics appears in the study of higher dimensional BMS invariant field theories. The study of BMSFTs is a very nascent activity as of now, and a lot of corners has not really been explored yet. Although works have appeared studying classical symmetries of higher dimensional BMS scalar fields \cite{Gupta:2020dtl, Bagchi:2022emh }, the systematic quantization and related vacuum structure of such theories are still mysterious. Since the transformation between ground states of $2d$ free scalar CFT and a BMS$_3$ invariant free scalar field offers such an unique connection, one could hope that similar structures also work out in higher dimensions, but concrete proposals are yet to materialize. One related state independent approach in this regard would be to use field theoretic techniques centered around symmetry algebras for BMS invariant theories, after modifying the approach for CFTs widely discussed in recent times \cite{Caputa:2018kdj, Erdmenger:2021wzc,Aip:2018qfv}.

Another interesting thing to note is the authors of \cite{Bagchi:2021qfe} found clear Lyapunovian behaviour in studying chaos for Carrollian conformal field theories in two dimensions. The situation there does not pertain to a transition from a relativistic CFT, but, intriguingly, actual intrinsic Carrollian dynamics does produce a chaotic spectrum. One may want to investigate the contradiction between these two approaches and learn more about such theories. We can also conjecture that something exciting is happening if one can make the parameter $\e$ purely imaginary in a particular setting and perhaps compute the Lyapunov index, but we would come back to these questions in a separate work.  
\section*{Acknowledgements}
The authors would like to thank Arjun Bagchi for comments on an earlier version of this manuscript.
The work of ArB is supported by the Quantum Gravity Unit of the Okinawa Institute of Science and Technology Graduate University (OIST). A.B. is supported by Start-Up Research Grant (SRG/2020/001380), Mathematical Research Impact Centric Support Grant (MTR/2021/000490) by the Department of Science and Technology Science and Engineering Research Board (India) and Relevant Research Project grant (58/14/12/2021- BRNS) by the Board Of Research In Nuclear Sciences (BRNS), Department of Atomic Energy, India.

\appendix

\bibliographystyle{utphysmodb}
\bibliography{ref.bib}

\providecommand{\href}[2]{#2}\begingroup\raggedright\begin{thebibliography}{100}

\bibitem{Maldacena:1997re}
J.~M. Maldacena,  {\em {The Large N limit of superconformal field theories and
  supergravity}}, Int. J. Theor. Phys. {\bf 38} (1999) 1113--1133
  [\href{http://www.arXiv.org/abs/hep-th/9711200}{{\tt hep-th/9711200}}],
[Adv. Theor. Math. Phys.2,231(1998)].

\bibitem{Bondi:1}
H.~Bondi, M.~G.~J. van~der Burg and A.~W.~K. Metzner,  {\em {Gravitational
  waves in general relativity. 7. Waves from axisymmetric isolated systems}},
  Proc. Roy. Soc. Lond. {\bf A269} (1962)
21--52.

\bibitem{Sachs:1962zza}
R.~Sachs,  {\em {Asymptotic symmetries in gravitational theory}}, Phys. Rev.
  {\bf 128} (1962)
2851--2864.

\bibitem{Bagchi:2012yk}
A.~Bagchi, S.~Detournay and D.~Grumiller,  {\em {Flat-Space Chiral Gravity}},
  Phys. Rev. Lett. {\bf 109} (2012) 151301
  [\href{http://www.arXiv.org/abs/1208.1658}{{\tt 1208.1658}}].

\bibitem{Bagchi:2012xr}
A.~Bagchi, S.~Detournay, R.~Fareghbal and J.~Sim{\'o}n,  {\em {Holography of 3D
  Flat Cosmological Horizons}}, Phys. Rev. Lett. {\bf 110} (2013), no.~14,
  141302
[\href{http://www.arXiv.org/abs/1208.4372}{{\tt 1208.4372}}].

\bibitem{Barnich:2012xq}
G.~Barnich,  {\em {Entropy of three-dimensional asymptotically flat
  cosmological solutions}}, JHEP {\bf 10} (2012) 095
[\href{http://www.arXiv.org/abs/1208.4371}{{\tt 1208.4371}}].

\bibitem{Bagchi:2013qva}
A.~Bagchi and R.~Basu,  {\em {3D Flat Holography: Entropy and Logarithmic
  Corrections}}, JHEP {\bf 03} (2014) 020
  [\href{http://www.arXiv.org/abs/1312.5748}{{\tt 1312.5748}}].

\bibitem{Bagchi:2015wna}
A.~Bagchi, D.~Grumiller and W.~Merbis,  {\em {Stress tensor correlators in
  three-dimensional gravity}}, Phys. Rev. D {\bf 93} (2016), no.~6, 061502
  [\href{http://www.arXiv.org/abs/1507.05620}{{\tt 1507.05620}}].

\bibitem{Bagchi:2014iea}
A.~Bagchi, R.~Basu, D.~Grumiller and M.~Riegler,  {\em {Entanglement entropy in
  Galilean conformal field theories and flat holography}}, Phys. Rev. Lett.
  {\bf 114} (2015), no.~11, 111602
[\href{http://www.arXiv.org/abs/1410.4089}{{\tt 1410.4089}}].

\bibitem{Jiang:2017ecm}
H.~Jiang, W.~Song and Q.~Wen,  {\em {Entanglement Entropy in Flat Holography}},
  JHEP {\bf 07} (2017) 142 [\href{http://www.arXiv.org/abs/1706.07552}{{\tt
  1706.07552}}].

\bibitem{Hijano:2017eii}
E.~Hijano and C.~Rabideau,  {\em {Holographic entanglement and Poincar\'e
  blocks in three-dimensional flat space}}, JHEP {\bf 05} (2018) 068
  [\href{http://www.arXiv.org/abs/1712.07131}{{\tt 1712.07131}}].

\bibitem{Bagchi:2013lma}
A.~Bagchi, S.~Detournay, D.~Grumiller and J.~Simon,  {\em {Cosmic Evolution
  from Phase Transition of Three-Dimensional Flat Space}}, Phys. Rev. Lett.
  {\bf 111} (2013), no.~18, 181301
[\href{http://www.arXiv.org/abs/1305.2919}{{\tt 1305.2919}}].

\bibitem{Detournay:2014fva}
S.~Detournay, D.~Grumiller, F.~Sch\"oller and J.~Sim\'on,  {\em {Variational
  principle and one-point functions in three-dimensional flat space Einstein
  gravity}}, Phys. Rev. D {\bf 89} (2014), no.~8, 084061
  [\href{http://www.arXiv.org/abs/1402.3687}{{\tt 1402.3687}}].

\bibitem{Hartong:2015usd}
J.~Hartong,  {\em {Holographic Reconstruction of 3D Flat Space-Time}}, JHEP
  {\bf 10} (2016) 104
[\href{http://www.arXiv.org/abs/1511.01387}{{\tt 1511.01387}}].

\bibitem{Hartong:2015xda}
J.~Hartong,  {\em {Gauging the Carroll Algebra and Ultra-Relativistic
  Gravity}}, JHEP {\bf 08} (2015) 069
  [\href{http://www.arXiv.org/abs/1505.05011}{{\tt 1505.05011}}].

\bibitem{Bagchi:2016geg}
A.~Bagchi, M.~Gary and Zodinmawia,  {\em {Bondi-Metzner-Sachs bootstrap}},
  Phys. Rev. {\bf D96} (2017), no.~2, 025007
[\href{http://www.arXiv.org/abs/1612.01730}{{\tt 1612.01730}}].

\bibitem{Barnich:2014cwa}
G.~Barnich, L.~Donnay, J.~Matulich and R.~Troncoso,  {\em {Asymptotic
  symmetries and dynamics of three-dimensional flat supergravity}}, JHEP {\bf
  08} (2014) 071
[\href{http://www.arXiv.org/abs/1407.4275}{{\tt 1407.4275}}].

\bibitem{Fareghbal:2014qga}
R.~Fareghbal and A.~Naseh,  {\em {Aspects of Flat/CCFT Correspondence}}, Class.
  Quant. Grav. {\bf 32} (2015) 135013
[\href{http://www.arXiv.org/abs/1408.6932}{{\tt 1408.6932}}].

\bibitem{Grumiller:2019xna}
D.~Grumiller, P.~Parekh and M.~Riegler,  {\em {Local quantum energy conditions
  in non-Lorentz-invariant quantum field theories}},
\href{http://www.arXiv.org/abs/1907.06650}{{\tt 1907.06650}}.

\bibitem{Ciambelli:2018wre}
L.~Ciambelli, C.~Marteau, A.~C. Petkou, P.~M. Petropoulos and K.~Siampos,  {\em
  {Flat holography and Carrollian fluids}}, JHEP {\bf 07} (2018) 165
  [\href{http://www.arXiv.org/abs/1802.06809}{{\tt 1802.06809}}].

\bibitem{Strominger:2017zoo}
A.~Strominger,  {\em {Lectures on the Infrared Structure of Gravity and Gauge
  Theory}},
\href{http://www.arXiv.org/abs/1703.05448}{{\tt 1703.05448}}.

\bibitem{Pasterski:2021rjz}
S.~Pasterski,  {\em {Lectures on celestial amplitudes}}, Eur. Phys. J. C {\bf
  81} (2021), no.~12, 1062 [\href{http://www.arXiv.org/abs/2108.04801}{{\tt
  2108.04801}}].

\bibitem{Raclariu:2021zjz}
A.-M. Raclariu,  {\em {Lectures on Celestial Holography}},
  \href{http://www.arXiv.org/abs/2107.02075}{{\tt 2107.02075}}.

\bibitem{Bagchi:2022emh}
A.~Bagchi, S.~Banerjee, R.~Basu and S.~Dutta,  {\em {Scattering Amplitudes:
  Celestial and Carrollian}}, \href{http://www.arXiv.org/abs/2202.08438}{{\tt
  2202.08438}}.

\bibitem{Donnay:2022aba}
L.~Donnay, A.~Fiorucci, Y.~Herfray and R.~Ruzziconi,  {\em {A Carrollian
  Perspective on Celestial Holography}},
  \href{http://www.arXiv.org/abs/2202.04702}{{\tt 2202.04702}}.

\bibitem{Bagchi:2009my}
A.~Bagchi and R.~Gopakumar,  {\em {Galilean Conformal Algebras and AdS/CFT}},
  JHEP {\bf 07} (2009) 037
[\href{http://www.arXiv.org/abs/0902.1385}{{\tt 0902.1385}}].

\bibitem{Duval_2014}
C.~Duval, G.~W. Gibbons and P.~A. Horvathy,  {\em Conformal Carroll groups and
  {BMS} symmetry}, Classical and Quantum Gravity {\bf 31} (apr, 2014) 092001.

\bibitem{Duval:2014uoa}
C.~Duval, G.~W. Gibbons, P.~A. Horvathy and P.~M. Zhang,  {\em {Carroll versus
  Newton and Galilei: two dual non-Einsteinian concepts of time}}, Class.
  Quant. Grav. {\bf 31} (2014) 085016
[\href{http://www.arXiv.org/abs/1402.0657}{{\tt 1402.0657}}].

\bibitem{Duval:2014uva}
C.~Duval, G.~W. Gibbons and P.~A. Horvathy,  {\em {Conformal Carroll groups and
  BMS symmetry}}, Class. Quant. Grav. {\bf 31} (2014) 092001
[\href{http://www.arXiv.org/abs/1402.5894}{{\tt 1402.5894}}].

\bibitem{Barnich:2010eb}
G.~Barnich and C.~Troessaert,  {\em {Aspects of the BMS/CFT correspondence}},
  JHEP {\bf 05} (2010) 062
[\href{http://www.arXiv.org/abs/1001.1541}{{\tt 1001.1541}}].

\bibitem{Bagchi:2010zz}
A.~Bagchi,  {\em {Correspondence between Asymptotically Flat Spacetimes and
  Nonrelativistic Conformal Field Theories}}, Phys. Rev. Lett. {\bf 105} (2010)
  171601 [\href{http://www.arXiv.org/abs/1006.3354}{{\tt 1006.3354}}].

\bibitem{Bagchi:2012cy}
A.~Bagchi and R.~Fareghbal,  {\em {BMS/GCA Redux: Towards Flatspace Holography
  from Non-Relativistic Symmetries}}, JHEP {\bf 10} (2012) 092
[\href{http://www.arXiv.org/abs/1203.5795}{{\tt 1203.5795}}].

\bibitem{Ciambelli:2018ojf}
L.~Ciambelli and C.~Marteau,  {\em {Carrollian conservation laws and Ricci-flat
  gravity}}, Class. Quant. Grav. {\bf 36} (2019), no.~8, 085004
  [\href{http://www.arXiv.org/abs/1810.11037}{{\tt 1810.11037}}].

\bibitem{Gupta:2020dtl}
N.~Gupta and N.~V. Suryanarayana,  {\em {Constructing Carrollian CFTs}}, JHEP
  {\bf 03} (2021) 194 [\href{http://www.arXiv.org/abs/2001.03056}{{\tt
  2001.03056}}].

\bibitem{Bagchi:2022eav}
A.~Bagchi, A.~Banerjee, S.~Dutta, K.~S. Kolekar and P.~Sharma,  {\em {Carroll
  covariant scalar fields in two dimensions}},
  \href{http://www.arXiv.org/abs/2203.13197}{{\tt 2203.13197}}.

\bibitem{Donnay:2019jiz}
L.~Donnay and C.~Marteau,  {\em {Carrollian Physics at the Black Hole
  Horizon}}, Class. Quant. Grav. {\bf 36} (2019), no.~16, 165002
  [\href{http://www.arXiv.org/abs/1903.09654}{{\tt 1903.09654}}].

\bibitem{Bagchi:2021qfe}
A.~Bagchi, S.~Chakrabortty, D.~Grumiller, B.~Radhakrishnan, M.~Riegler and
  A.~Sinha,  {\em {Non-Lorentzian chaos and cosmological holography}}, Phys.
  Rev. D {\bf 104} (2021), no.~10, L101901
  [\href{http://www.arXiv.org/abs/2106.07649}{{\tt 2106.07649}}].

\bibitem{Schild:1976vq}
A.~Schild,  {\em {Classical Null Strings}}, Phys. Rev. {\bf D16} (1977)
1722.

\bibitem{Isberg:1993av}
J.~Isberg, U.~Lindstrom, B.~Sundborg and G.~Theodoridis,  {\em {Classical and
  quantized tensionless strings}}, Nucl. Phys. {\bf B411} (1994) 122--156
[\href{http://www.arXiv.org/abs/hep-th/9307108}{{\tt hep-th/9307108}}].

\bibitem{Bagchi:2013bga}
A.~Bagchi,  {\em {Tensionless Strings and Galilean Conformal Algebra}}, JHEP
  {\bf 05} (2013) 141
[\href{http://www.arXiv.org/abs/1303.0291}{{\tt 1303.0291}}].

\bibitem{Bagchi:2015nca}
A.~Bagchi, S.~Chakrabortty and P.~Parekh,  {\em {Tensionless Strings from
  Worldsheet Symmetries}}, JHEP {\bf 01} (2016) 158
[\href{http://www.arXiv.org/abs/1507.04361}{{\tt 1507.04361}}].

\bibitem{Hao:2021urq}
P.-x. Hao, W.~Song, X.~Xie and Y.~Zhong,  {\em {A BMS-invariant free scalar
  model}}, \href{http://www.arXiv.org/abs/2111.04701}{{\tt 2111.04701}}.

\bibitem{Rodriguez:2021tcz}
P.~Rodr\'\i{}guez, D.~Tempo and R.~Troncoso,  {\em {Mapping relativistic to
  ultra/non-relativistic conformal symmetries in 2D and finite $
  \sqrt{T\overline{T}} $ deformations}}, JHEP {\bf 11} (2021) 133
  [\href{http://www.arXiv.org/abs/2106.09750}{{\tt 2106.09750}}].

\bibitem{Bagchi:2022nvj}
A.~Bagchi, A.~Banerjee and H.~Muraki,  {\em {Boosting to BMS}},
  \href{http://www.arXiv.org/abs/2205.05094}{{\tt 2205.05094}}.

\bibitem{Stanford:2014jda}
D.~Stanford and L.~Susskind,  {\em {Complexity and Shock Wave Geometries}},
  Phys. Rev. D {\bf 90} (2014), no.~12, 126007
  [\href{http://www.arXiv.org/abs/1406.2678}{{\tt 1406.2678}}].

\bibitem{Brown:2015bva}
A.~R. Brown, D.~A. Roberts, L.~Susskind, B.~Swingle and Y.~Zhao,  {\em
  {Holographic Complexity Equals Bulk Action?}}, Phys. Rev. Lett. {\bf 116}
  (2016), no.~19, 191301 [\href{http://www.arXiv.org/abs/1509.07876}{{\tt
  1509.07876}}].

\bibitem{Jefferson1}
R.~Jefferson and R.~C. Myers,  {\em {Circuit complexity in quantum field
  theory}}, JHEP {\bf 10} (2017) 107
[\href{http://www.arXiv.org/abs/1707.08570}{{\tt 1707.08570}}].

\bibitem{Chapman:2017rqy}
S.~Chapman, M.~P. Heller, H.~Marrochio and F.~Pastawski,  {\em {Toward a
  Definition of Complexity for Quantum Field Theory States}}, Phys. Rev. Lett.
  {\bf 120} (2018), no.~12, 121602
  [\href{http://www.arXiv.org/abs/1707.08582}{{\tt 1707.08582}}].

\bibitem{Caputa:2017yrh}
P.~Caputa, N.~Kundu, M.~Miyaji, T.~Takayanagi and K.~Watanabe,  {\em {Liouville
  Action as Path-Integral Complexity: From Continuous Tensor Networks to
  AdS/CFT}}, JHEP {\bf 11} (2017) 097
  [\href{http://www.arXiv.org/abs/1706.07056}{{\tt 1706.07056}}].

\bibitem{me1}
T.~Ali, A.~Bhattacharyya, S.~Shajidul~Haque, E.~H. Kim and N.~Moynihan,  {\em
  {Time Evolution of Complexity: A Critique of Three Methods}}, JHEP {\bf 04}
  (2019) 087 [\href{http://www.arXiv.org/abs/1810.02734}{{\tt 1810.02734}}].

\bibitem{Bhattacharyya:2018bbv}
A.~Bhattacharyya, A.~Shekar and A.~Sinha,  {\em {Circuit complexity in
  interacting QFTs and RG flows}}, JHEP {\bf 10} (2018) 140
  [\href{http://www.arXiv.org/abs/1808.03105}{{\tt 1808.03105}}].

\bibitem{Hackl:2018ptj}
L.~Hackl and R.~C. Myers,  {\em {Circuit complexity for free fermions}}, JHEP
  {\bf 07} (2018) 139 [\href{http://www.arXiv.org/abs/1803.10638}{{\tt
  1803.10638}}].

\bibitem{Khan:2018rzm}
R.~Khan, C.~Krishnan and S.~Sharma,  {\em {Circuit Complexity in Fermionic
  Field Theory}}, Phys. Rev. D {\bf 98} (2018), no.~12, 126001
  [\href{http://www.arXiv.org/abs/1801.07620}{{\tt 1801.07620}}].

\bibitem{Alves:2018qfv}
D.~W.~F. Alves and G.~Camilo,  {\em {Evolution of complexity following a
  quantum quench in free field theory}}, JHEP {\bf 06} (2018) 029
  [\href{http://www.arXiv.org/abs/1804.00107}{{\tt 1804.00107}}].

\bibitem{Camargo:2018eof}
H.~A. Camargo, P.~Caputa, D.~Das, M.~P. Heller and R.~Jefferson,  {\em
  {Complexity as a novel probe of quantum quenches: universal scalings and
  purifications}}, Phys. Rev. Lett. {\bf 122} (2019), no.~8, 081601
  [\href{http://www.arXiv.org/abs/1807.07075}{{\tt 1807.07075}}].

\bibitem{Ali:2018aon}
T.~Ali, A.~Bhattacharyya, S.~Shajidul~Haque, E.~H. Kim and N.~Moynihan,  {\em
  {Post-Quench Evolution of Complexity and Entanglement in a Topological
  System}}, Phys. Lett. B {\bf 811} (2020) 135919
  [\href{http://www.arXiv.org/abs/1811.05985}{{\tt 1811.05985}}].

\bibitem{Bhattacharyya:2018wym}
A.~Bhattacharyya, P.~Caputa, S.~R. Das, N.~Kundu, M.~Miyaji and T.~Takayanagi,
  {\em {Path-Integral Complexity for Perturbed CFTs}}, JHEP {\bf 07} (2018) 086
  [\href{http://www.arXiv.org/abs/1804.01999}{{\tt 1804.01999}}].

\bibitem{Caputa:2018kdj}
P.~Caputa and J.~M. Magan,  {\em {Quantum Computation as Gravity}}, Phys. Rev.
  Lett. {\bf 122} (2019), no.~23, 231302
  [\href{http://www.arXiv.org/abs/1807.04422}{{\tt 1807.04422}}].

\bibitem{Bhattacharyya:2019kvj}
A.~Bhattacharyya, P.~Nandy and A.~Sinha,  {\em {Renormalized Circuit
  Complexity}}, Phys. Rev. Lett. {\bf 124} (2020), no.~10, 101602
[\href{http://www.arXiv.org/abs/1907.08223}{{\tt 1907.08223}}].

\bibitem{Flory:2020eot}
M.~Flory and M.~P. Heller,  {\em {Geometry of Complexity in Conformal Field
  Theory}}, Phys. Rev. Res. {\bf 2} (2020), no.~4, 043438
  [\href{http://www.arXiv.org/abs/2005.02415}{{\tt 2005.02415}}].

\bibitem{Erdmenger:2020sup}
J.~Erdmenger, M.~Gerbershagen and A.-L. Weigel,  {\em {Complexity measures from
  geometric actions on Virasoro and Kac-Moody orbits}}, JHEP {\bf 11} (2020)
  003 [\href{http://www.arXiv.org/abs/2004.03619}{{\tt 2004.03619}}].

\bibitem{cosmology1}
A.~Bhattacharyya, S.~Das, S.~S. Haque and B.~Underwood,  {\em Cosmological
  complexity}, Physical Review D {\bf 101} (May, 2020).

\bibitem{cosmology2}
A.~Bhattacharyya, S.~Das, S.~S. Haque and B.~Underwood,  {\em Rise of
  cosmological complexity: Saturation of growth and chaos}, Physical Review
  Research {\bf 2} (Aug, 2020).

\bibitem{DiGiulio:2020hlz}
G.~Di~Giulio and E.~Tonni,  {\em {Complexity of mixed Gaussian states from
  Fisher information geometry}}, JHEP {\bf 12} (2020) 101
  [\href{http://www.arXiv.org/abs/2006.00921}{{\tt 2006.00921}}].

\bibitem{Caceres:2019pgf}
E.~Caceres, S.~Chapman, J.~D. Couch, J.~P. Hernandez, R.~C. Myers and S.-M.
  Ruan,  {\em {Complexity of Mixed States in QFT and Holography}}, JHEP {\bf
  03} (2020) 012 [\href{http://www.arXiv.org/abs/1909.10557}{{\tt
  1909.10557}}].

\bibitem{Chen:2020nlj}
B.~Chen, B.~Czech and Z.-z. Wang,  {\em {Query complexity and cutoff dependence
  of the CFT2 ground state}}, Phys. Rev. D {\bf 103} (2021), no.~2, 026015
  [\href{http://www.arXiv.org/abs/2004.11377}{{\tt 2004.11377}}].

\bibitem{Czech:2017ryf}
B.~Czech,  {\em {Einstein Equations from Varying Complexity}}, Phys. Rev. Lett.
  {\bf 120} (2018), no.~3, 031601
  [\href{http://www.arXiv.org/abs/1706.00965}{{\tt 1706.00965}}].

\bibitem{Camargo:2019isp}
H.~A. Camargo, M.~P. Heller, R.~Jefferson and J.~Knaute,  {\em {Path integral
  optimization as circuit complexity}}, Phys. Rev. Lett. {\bf 123} (2019),
  no.~1, 011601 [\href{http://www.arXiv.org/abs/1904.02713}{{\tt 1904.02713}}].

\bibitem{Chapman:2018hou}
S.~Chapman, J.~Eisert, L.~Hackl, M.~P. Heller, R.~Jefferson, H.~Marrochio and
  R.~C. Myers,  {\em {Complexity and entanglement for thermofield double
  states}}, SciPost Phys. {\bf 6} (2019), no.~3, 034
  [\href{http://www.arXiv.org/abs/1810.05151}{{\tt 1810.05151}}].

\bibitem{Doroudiani:2019llj}
M.~Doroudiani, A.~Naseh and R.~Pirmoradian,  {\em {Complexity for Charged
  Thermofield Double States}}, JHEP {\bf 01} (2020) 120
  [\href{http://www.arXiv.org/abs/1910.08806}{{\tt 1910.08806}}].

\bibitem{Geng:2019yxo}
H.~Geng,  {\em {$T\bar{T}$ Deformation and the Complexity=Volume Conjecture}},
  Fortsch. Phys. {\bf 68} (2020), no.~7, 2000036
  [\href{http://www.arXiv.org/abs/1910.08082}{{\tt 1910.08082}}].

\bibitem{Couch:2021wsm}
J.~Couch, Y.~Fan and S.~Shashi,  {\em {Circuit Complexity in Topological
  Quantum Field Theory}}, \href{http://www.arXiv.org/abs/2108.13427}{{\tt
  2108.13427}}.

\bibitem{Bhattacharyya:2021fii}
A.~Bhattacharyya, T.~Hanif, S.~S. Haque and M.~K. Rahman,  {\em {Complexity for
  an open quantum system}}, Phys. Rev. D {\bf 105} (2022), no.~4, 046011
  [\href{http://www.arXiv.org/abs/2112.03955}{{\tt 2112.03955}}].

\bibitem{Bhattacharyya:2019txx}
A.~Bhattacharyya, W.~Chemissany, S.~Shajidul~Haque and B.~Yan,  {\em {Towards
  the web of quantum chaos diagnostics}}, Eur. Phys. J. C {\bf 82} (2022),
  no.~1, 87 [\href{http://www.arXiv.org/abs/1909.01894}{{\tt 1909.01894}}].

\bibitem{Erdmenger:2021wzc}
J.~Erdmenger, M.~Flory, M.~Gerbershagen, M.~P. Heller and A.-L. Weigel,  {\em
  {Exact Gravity Duals for Simple Quantum Circuits}},
  \href{http://www.arXiv.org/abs/2112.12158}{{\tt 2112.12158}}.

\bibitem{Chagnet:2021uvi}
N.~Chagnet, S.~Chapman, J.~de~Boer and C.~Zukowski,  {\em {Complexity for
  Conformal Field Theories in General Dimensions}}, Phys. Rev. Lett. {\bf 128}
  (2022), no.~5, 051601 [\href{http://www.arXiv.org/abs/2103.06920}{{\tt
  2103.06920}}].

\bibitem{Bhattacharyya:2022ren}
A.~Bhattacharyya, G.~Katoch and S.~R. Roy,  {\em {Complexity of warped
  conformal field theory}}, \href{http://www.arXiv.org/abs/2202.09350}{{\tt
  2202.09350}}.

\bibitem{Chapman:2021jbh}
S.~Chapman and G.~Policastro,  {\em {Quantum computational complexity from
  quantum information to black holes and back}}, Eur. Phys. J. C {\bf 82}
  (2022), no.~2, 128 [\href{http://www.arXiv.org/abs/2110.14672}{{\tt
  2110.14672}}].

\bibitem{Bhattacharyya:2021cwf}
A.~Bhattacharyya,  {\em {Circuit complexity and (some of) its applications}},
  Int. J. Mod. Phys. E {\bf 30} (2021), no.~07, 2130005.

\bibitem{Koch:2021tvp}
R.~d.~M. Koch, M.~Kim and H.~J.~R. Van~Zyl,  {\em {Complexity from spinning
  primaries}}, JHEP {\bf 12} (2021) 030
  [\href{http://www.arXiv.org/abs/2108.10669}{{\tt 2108.10669}}].

\bibitem{Couch:2018phr}
J.~Couch, S.~Eccles, T.~Jacobson and P.~Nguyen,  {\em {Holographic Complexity
  and Volume}}, JHEP {\bf 11} (2018) 044
  [\href{http://www.arXiv.org/abs/1807.02186}{{\tt 1807.02186}}].

\bibitem{Belin:2021bga}
A.~Belin, R.~C. Myers, S.-M. Ruan, G.~S\'arosi and A.~J. Speranza,  {\em {Does
  Complexity Equal Anything?}}, Phys. Rev. Lett. {\bf 128} (2022), no.~8,
  081602 [\href{http://www.arXiv.org/abs/2111.02429}{{\tt 2111.02429}}].

\bibitem{Agon:2018zso}
C.~A. Ag\'on, M.~Headrick and B.~Swingle,  {\em {Subsystem Complexity and
  Holography}}, JHEP {\bf 02} (2019) 145
  [\href{http://www.arXiv.org/abs/1804.01561}{{\tt 1804.01561}}].

\bibitem{Belin:2018bpg}
A.~Belin, A.~Lewkowycz and G.~S\'arosi,  {\em {Complexity and the bulk volume,
  a new York time story}}, JHEP {\bf 03} (2019) 044
  [\href{http://www.arXiv.org/abs/1811.03097}{{\tt 1811.03097}}].

\bibitem{NHJ}
N.~R. Hunter-Jones,  {\em {Chaos and Randomness in Strongly-Interacting Quantum
  Systems.}}, Dissertation (Ph.D.), California Institute of Technology. {\bf
  52} (2018) 1--4.

\bibitem{Jahnke:2018off}
V.~Jahnke,  {\em {Recent developments in the holographic description of quantum
  chaos}}, Adv. High Energy Phys. {\bf 2019} (2019) 9632708
  [\href{http://www.arXiv.org/abs/1811.06949}{{\tt 1811.06949}}].

\bibitem{Kitaev2015}
A.~Kitaev,  {\em A simple model of quantum holography}, Proceedings of the KITP
  Program: Entanglement in Strongly-Correlated Quantum Matter, (Kavli Institute
  for Theoretical Physics, Santa Barbara) {\bf Vol. 7} (2015).

\bibitem{Larkin1969}
A.~I. {Larkin} and Y.~N. {Ovchinnikov},  {\em {Quasiclassical Method in the
  Theory of Superconductivity}}, Soviet Journal of Experimental and Theoretical
  Physics {\bf 28} (June, 1969) 1200.

\bibitem{Maldacena_2016}
J.~Maldacena, S.~H. Shenker and D.~Stanford,  {\em A bound on chaos}, Journal
  of High Energy Physics {\bf 2016} (Aug, 2016).

\bibitem{Hashimoto:2017oit}
K.~Hashimoto, K.~Murata and R.~Yoshii,  {\em {Out-of-time-order correlators in
  quantum mechanics}}, JHEP {\bf 10} (2017) 138
  [\href{http://www.arXiv.org/abs/1703.09435}{{\tt 1703.09435}}].

\bibitem{Parker:2018yvk}
D.~E. Parker, X.~Cao, A.~Avdoshkin, T.~Scaffidi and E.~Altman,  {\em {A
  Universal Operator Growth Hypothesis}}, Phys. Rev. X {\bf 9} (2019), no.~4,
  041017 [\href{http://www.arXiv.org/abs/1812.08657}{{\tt 1812.08657}}].

\bibitem{Dymarsky:2021bjq}
A.~Dymarsky and M.~Smolkin,  {\em {Krylov complexity in conformal field
  theory}}, Phys. Rev. D {\bf 104} (2021), no.~8, L081702
  [\href{http://www.arXiv.org/abs/2104.09514}{{\tt 2104.09514}}].

\bibitem{Barbon:2019wsy}
J.~L.~F. Barb\'on, E.~Rabinovici, R.~Shir and R.~Sinha,  {\em {On The Evolution
  Of Operator Complexity Beyond Scrambling}}, JHEP {\bf 10} (2019) 264
  [\href{http://www.arXiv.org/abs/1907.05393}{{\tt 1907.05393}}].

\bibitem{Rabinovici:2020ryf}
E.~Rabinovici, A.~S\'anchez-Garrido, R.~Shir and J.~Sonner,  {\em {Operator
  complexity: a journey to the edge of Krylov space}}, JHEP {\bf 06} (2021) 062
  [\href{http://www.arXiv.org/abs/2009.01862}{{\tt 2009.01862}}].

\bibitem{Kar:2021nbm}
A.~Kar, L.~Lamprou, M.~Rozali and J.~Sully,  {\em {Random matrix theory for
  complexity growth and black hole interiors}}, JHEP {\bf 01} (2022) 016
  [\href{http://www.arXiv.org/abs/2106.02046}{{\tt 2106.02046}}].

\bibitem{Caputa:2021sib}
P.~Caputa, J.~M. Magan and D.~Patramanis,  {\em {Geometry of Krylov
  complexity}}, Phys. Rev. Res. {\bf 4} (2022), no.~1, 013041
  [\href{http://www.arXiv.org/abs/2109.03824}{{\tt 2109.03824}}].

\bibitem{Balasubramanian:2022tpr}
V.~Balasubramanian, P.~Caputa, J.~Magan and Q.~Wu,  {\em {Quantum chaos and the
  complexity of spread of states}},
  \href{http://www.arXiv.org/abs/2202.06957}{{\tt 2202.06957}}.

\bibitem{Bhattacharjee:2022vlt}
B.~Bhattacharjee, X.~Cao, P.~Nandy and T.~Pathak,  {\em {Krylov complexity in
  saddle-dominated scrambling}},
  \href{http://www.arXiv.org/abs/2203.03534}{{\tt 2203.03534}}.

\bibitem{Muck:2022xfc}
W.~M\"uck and Y.~Yang,  {\em {Krylov complexity and orthogonal polynomials}},
  \href{http://www.arXiv.org/abs/2205.12815}{{\tt 2205.12815}}.

\bibitem{Bagchi:2020fpr}
A.~Bagchi, A.~Banerjee, S.~Chakrabortty, S.~Dutta and P.~Parekh,  {\em {A tale
  of three \textemdash{} tensionless strings and vacuum structure}}, JHEP {\bf
  04} (2020) 061 [\href{http://www.arXiv.org/abs/2001.00354}{{\tt
  2001.00354}}].

\bibitem{Bagchi:2021ban}
A.~Bagchi, A.~Banerjee, S.~Chakrabortty and R.~Chatterjee,  {\em {A Rindler
  Road to Carrollian Worldsheets}},
  \href{http://www.arXiv.org/abs/2111.01172}{{\tt 2111.01172}}.

\bibitem{NL1}
M.~A. Nielsen, M.~R. Dowling, M.~Gu and A.~C. Doherty,  {\em Quantum
  computation as geometry}, Science {\bf 311} (Feb., 2006) 1133--1135.

\bibitem{NL2}
M.~R. Nielsen, M.~A.and~Dowling,  {\em {The geometry of quantum computation}},
  Science {\bf 311} (2006), no.~4, 1133--1135
[\href{http://www.arXiv.org/abs/0701004}{{\tt 0701004}}].

\bibitem{NL3}
M.~A. Nielsen,  {\em {A geometric approach to quantum circuit lower bounds}},
  Science {\bf 311} (2006), no.~4, 92
[\href{http://www.arXiv.org/abs/0502070}{{\tt 0502070}}].

\bibitem{Guo:2018kzl}
M.~Guo, J.~Hernandez, R.~C. Myers and S.-M. Ruan,  {\em {Circuit Complexity for
  Coherent States}}, JHEP {\bf 10} (2018) 011
  [\href{http://www.arXiv.org/abs/1807.07677}{{\tt 1807.07677}}].

\bibitem{Camargo_2019}
H.~A. Camargo, P.~Caputa, D.~Das, M.~P. Heller and R.~Jefferson,  {\em
  Complexity as a Novel Probe of Quantum Quenches: Universal Scalings and
  Purifications}, Physical Review Letters {\bf 122} (Feb, 2019).

\bibitem{Liu:2019aji}
F.~Liu, S.~Whitsitt, J.~B. Curtis, R.~Lundgren, P.~Titum, Z.-C. Yang, J.~R.
  Garrison and A.~V. Gorshkov,  {\em {Circuit complexity across a topological
  phase transition}}, Phys. Rev. Res. {\bf 2} (2020), no.~1, 013323
  [\href{http://www.arXiv.org/abs/1902.10720}{{\tt 1902.10720}}].

\bibitem{Bagchi:2019cay}
A.~Bagchi, A.~Banerjee and P.~Parekh,  {\em {Tensionless Path from Closed to
  Open Strings}}, Phys. Rev. Lett. {\bf 123} (2019), no.~11, 111601
[\href{http://www.arXiv.org/abs/1905.11732}{{\tt 1905.11732}}].

\bibitem{Provost:1980nc}
J.~P. Provost and G.~Vallee,  {\em {Riemannian Structure on manifolds of
  quantum states}}, Commun. Math. Phys. {\bf 76} (1980) 289--301.

\bibitem{Berry:1984jv}
M.~V. Berry,  {\em {Quantal phase factors accompanying adiabatic changes}},
  Proc. Roy. Soc. Lond. A {\bf 392} (1984) 45--57.

\bibitem{Nogueira:2021ngh}
F.~S. Nogueira, S.~Banerjee, M.~Dorband, R.~Meyer, J.~v.~d. Brink and
  J.~Erdmenger,  {\em {Geometric phases distinguish entangled states in
  wormhole quantum mechanics}}, Phys. Rev. D {\bf 105} (2022), no.~8, L081903
  [\href{http://www.arXiv.org/abs/2109.06190}{{\tt 2109.06190}}].

\bibitem{Gharibyan:2018fax}
H.~Gharibyan, M.~Hanada, B.~Swingle and M.~Tezuka,  {\em {Quantum Lyapunov
  Spectrum}}, JHEP {\bf 04} (2019) 082
  [\href{http://www.arXiv.org/abs/1809.01671}{{\tt 1809.01671}}].

\bibitem{Bhattacharyya:2020art}
A.~Bhattacharyya, W.~Chemissany, S.~S. Haque, J.~Murugan and B.~Yan,  {\em {The
  Multi-faceted Inverted Harmonic Oscillator: Chaos and Complexity}}, SciPost
  Phys. Core {\bf 4} (2021) 002
  [\href{http://www.arXiv.org/abs/2007.01232}{{\tt 2007.01232}}].

\bibitem{vonKeyserlingk:2017dyr}
C.~von Keyserlingk, T.~Rakovszky, F.~Pollmann and S.~Sondhi,  {\em {Operator
  hydrodynamics, OTOCs, and entanglement growth in systems without conservation
  laws}}, Phys. Rev. X {\bf 8} (2018), no.~2, 021013
  [\href{http://www.arXiv.org/abs/1705.08910}{{\tt 1705.08910}}].

\bibitem{Nahum:2017yvy}
A.~Nahum, S.~Vijay and J.~Haah,  {\em {Operator Spreading in Random Unitary
  Circuits}}, Phys. Rev. X {\bf 8} (2018), no.~2, 021014
  [\href{http://www.arXiv.org/abs/1705.08975}{{\tt 1705.08975}}].

\bibitem{Khemani:2017nda}
V.~Khemani, A.~Vishwanath and D.~A. Huse,  {\em {Operator spreading and the
  emergence of dissipation in unitary dynamics with conservation laws}}, Phys.
  Rev. X {\bf 8} (2018), no.~3, 031057
  [\href{http://www.arXiv.org/abs/1710.09835}{{\tt 1710.09835}}].

\bibitem{Chan:2017kzq}
A.~Chan, A.~De~Luca and J.~T. Chalker,  {\em {Solution of a minimal model for
  many-body quantum chaos}}, Phys. Rev. X {\bf 8} (2018), no.~4, 041019
  [\href{http://www.arXiv.org/abs/1712.06836}{{\tt 1712.06836}}].

\bibitem{Bagchi:2020ats}
A.~Bagchi, A.~Banerjee and S.~Chakrabortty,  {\em {Rindler Physics on the
  String Worldsheet}}, Phys. Rev. Lett. {\bf 126} (2021), no.~3, 031601
  [\href{http://www.arXiv.org/abs/2009.01408}{{\tt 2009.01408}}].

\bibitem{Ali:2019zcj}
T.~Ali, A.~Bhattacharyya, S.~S. Haque, E.~H. Kim, N.~Moynihan and J.~Murugan,
  {\em {Chaos and Complexity in Quantum Mechanics}}, Phys. Rev. D {\bf 101}
  (2020), no.~2, 026021 [\href{http://www.arXiv.org/abs/1905.13534}{{\tt
  1905.13534}}].

\bibitem{haqueReducedDensityMatrix}
A.~Bhattacharyya, S.~S. Haque and E.~H. Kim,  {\em {Complexity from the Reduced
  Density Matrix: a new Diagnostic for Chaos}}, JHEP {\bf 10} (2021) 028
  [\href{http://www.arXiv.org/abs/2011.04705}{{\tt 2011.04705}}].

\bibitem{Ryuchaos}
J.~Kudler-Flam, L.~Nie and S.~Ryu,  {\em {Conformal field theory and the web of
  quantum chaos diagnostics}}, JHEP {\bf 01} (2020) 175
  [\href{http://www.arXiv.org/abs/1910.14575}{{\tt 1910.14575}}].

\bibitem{Magan2018-ua}
J.~M. Mag{\'a}n,  {\em Black holes, complexity and quantum chaos}, Journal of
  High Energy Physics {\bf 2018} (Sept., 2018) 43.

\bibitem{Qu:2021ius}
L.-C. Qu, J.~Chen and Y.-X. Liu,  {\em {Chaos and Complexity for Inverted
  Harmonic Oscillators}}, \href{http://www.arXiv.org/abs/2111.07351}{{\tt
  2111.07351}}.

\bibitem{Balasubramanian:2019wgd}
V.~Balasubramanian, M.~Decross, A.~Kar and O.~Parrikar,  {\em {Quantum
  Complexity of Time Evolution with Chaotic Hamiltonians}}, JHEP {\bf 01}
  (2020) 134 [\href{http://www.arXiv.org/abs/1905.05765}{{\tt 1905.05765}}].

\bibitem{Yang:2019iav}
R.-Q. Yang and K.-Y. Kim,  {\em {Time evolution of the complexity in chaotic
  systems: a concrete example}}, JHEP {\bf 05} (2020) 045
  [\href{http://www.arXiv.org/abs/1906.02052}{{\tt 1906.02052}}].

\bibitem{Kuwahara:2020chn}
T.~Kuwahara and K.~Saito,  {\em {Absence of fast scrambling in
  thermodynamically stable long-range interacting systems}}, Phys. Rev. Lett.
  {\bf 126} (2021), no.~3, 030604
  [\href{http://www.arXiv.org/abs/2009.10124}{{\tt 2009.10124}}].

\bibitem{Aip:2018qfv}
A.~Bhattacharyya and P.~Nandi,  {\em {To appear}}

\end{thebibliography}\endgroup

\end{document}